\documentstyle[11pt,epsf]{article}
\begin{document}
\newcommand {\ee}{\end{equation}}
\newcommand {\bea}{\begin{eqnarray}}
\newcommand {\eea}{\end{eqnarray}}
\newcommand {\nn}{\nonumber \\}
\newcommand {\Tr}{{\rm Tr\,}}
\newcommand {\tr}{{\rm tr\,}}
\newcommand {\e}{{\rm e}}
\newcommand {\etal}{{\it et al.}}
\newcommand {\m}{\mu}
\newcommand {\n}{\nu}
\newcommand {\pl}{\partial}
\newcommand {\p} {\phi}
\newcommand {\vp}{\varphi}
\newcommand {\vpc}{\varphi_c}
\newcommand {\al}{\alpha}
\newcommand {\be}{\beta}
\newcommand {\ga}{\gamma}
\newcommand {\Ga}{\Gamma}
\newcommand {\x}{\xi}
\newcommand {\ka}{\kappa}
\newcommand {\la}{\lambda}
\newcommand {\La}{\Lambda}
\newcommand {\si}{\sigma}
\newcommand {\Si}{\Sigma}
\newcommand {\th}{\theta}
\newcommand {\Th}{\Theta}
\newcommand {\om}{\omega}
\newcommand {\Om}{\Omega}
\newcommand {\ep}{\epsilon}
\newcommand {\vep}{\varepsilon}
\newcommand {\na}{\nabla}
\newcommand {\del}  {\delta}
\newcommand {\Del}  {\Delta}
\newcommand {\mn}{{\mu\nu}}
\newcommand {\ls}   {{\lambda\sigma}}
\newcommand {\ab}   {{\alpha\beta}}
\newcommand {\half}{ {\frac{1}{2}} }
\newcommand {\third}{ {\frac{1}{3}} }
\newcommand {\fourth} {\frac{1}{4} }
\newcommand {\sixth} {\frac{1}{6} }
\newcommand {\sqg} {\sqrt{g}}
\newcommand {\fg}  {\sqrt[4]{g}}
\newcommand {\invfg}  {\frac{1}{\sqrt[4]{g}}}
\newcommand {\sqZ} {\sqrt{Z}}
\newcommand {\gbar}{\bar{g}}
\newcommand {\sqk} {\sqrt{\kappa}}
\newcommand {\sqt} {\sqrt{t}}
\newcommand {\reg} {\frac{1}{\epsilon}}
\newcommand {\fpisq} {(4\pi)^2}
\newcommand {\Lcal}{{\cal L}}
\newcommand {\Ocal}{{\cal O}}
\newcommand {\Dcal}{{\cal D}}
\newcommand {\Ncal}{{\cal N}}
\newcommand {\Mcal}{{\cal M}}
\newcommand {\scal}{{\cal s}}
\newcommand {\Dvec}{{\hat D}}   
\newcommand {\dvec}{{\vec d}}
\newcommand {\Evec}{{\vec E}}
\newcommand {\Hvec}{{\vec H}}
\newcommand {\Vvec}{{\vec V}}
\newcommand {\Btil}{{\tilde B}}
\newcommand {\ctil}{{\tilde c}}
\newcommand {\Ftil}{{\tilde F}}
\newcommand {\Stil}{{\tilde S}}
\newcommand {\Ztil}{{\tilde Z}}
\newcommand {\altil}{{\tilde \alpha}}
\newcommand {\betil}{{\tilde \beta}}
\newcommand {\latil}{{\tilde \lambda}}
\newcommand {\ptil}{{\tilde \phi}}
\newcommand {\Ptil}{{\tilde P}}
\newcommand {\natil} {{\tilde \nabla}}
\newcommand {\ttil} {{\tilde t}}
\newcommand {\Rhat}{{\hat R}}
\newcommand {\Shat}{{\hat S}}
\newcommand {\shat}{{\hat s}}
\newcommand {\Dhat}{{\hat D}}   
\newcommand {\Vhat}{{\hat V}}   
\newcommand {\xhat}{{\hat x}}
\newcommand {\Zhat}{{\hat Z}}
\newcommand {\Gahat}{{\hat \Gamma}}
\newcommand {\nah} {{\hat \nabla}}
\newcommand {\gh}  {{\hat g}}
\newcommand {\labar}{{\bar \lambda}}
\newcommand {\cbar}{{\bar c}}
\newcommand {\bbar}{{\bar b}}
\newcommand {\Bbar}{{\bar B}}
\newcommand {\psibar}{{\bar \psi}}
\newcommand {\chibar}{{\bar \chi}}
\newcommand {\bbartil}{{\tilde {\bar b}}}
\newcommand  {\vz}{{v_0}}
\newcommand  {\ez}{{e_0}}
\newcommand  {\rhc}{{\rho_c}}
\newcommand {\intfx} {{\int d^4x}}
\newcommand {\inttx} {{\int d^2x}}
\newcommand {\change} {\leftrightarrow}
\newcommand {\ra} {\rightarrow}
\newcommand {\larrow} {\leftarrow}
\newcommand {\ul}   {\underline}
\newcommand {\pr}   {{\quad .}}
\newcommand {\com}  {{\quad ,}}
\newcommand {\q}    {\quad}
\newcommand {\qq}   {\quad\quad}
\newcommand {\qqq}   {\quad\quad\quad}
\newcommand {\qqqq}   {\quad\quad\quad\quad}
\newcommand {\qqqqq}   {\quad\quad\quad\quad\quad}
\newcommand {\qqqqqq}   {\quad\quad\quad\quad\quad\quad}
\newcommand {\qqqqqqq}   {\quad\quad\quad\quad\quad\quad\quad}
\newcommand {\lb}    {\linebreak}
\newcommand {\nl}    {\newline}

\newcommand {\vs}[1]  { \vspace*{#1 cm} }

\newcommand {\MPL}  {Mod.Phys.Lett.}
\newcommand {\NP}   {Nucl.Phys.}
\newcommand {\PL}   {Phys.Lett.}
\newcommand {\PR}   {Phys.Rev.}
\newcommand {\PRL}   {Phys.Rev.Lett.}
\newcommand {\CMP}  {Commun.Math.Phys.}
\newcommand {\JMP}  {Jour.Math.Phys.}
\newcommand {\AP}   {Ann.of Phys.}
\newcommand {\PTP}  {Prog.Theor.Phys.}
\newcommand {\NC}   {Nuovo Cim.}
\newcommand {\CQG}  {Class.Quantum.Grav.}


\font\smallr=cmr5
\def\ocirc#1{#1^{^{{\hbox{\smallr\llap{o}}}}}}
\def\ogamma{\ocirc{\gamma}{}}
\def\oM{{\buildrel {\hbox{\smallr{o}}} \over M}}
\def\osigma{\ocirc{\sigma}{}}

\def\overleftrightarrow#1{\vbox{\ialign{##\crcr
 $\leftrightarrow$\crcr\noalign{\kern-1pt\nointerlineskip}
 $\hfil\displaystyle{#1}\hfil$\crcr}}}
\def\overnab{{\overleftrightarrow\nabslash}}

\def\va{{a}}
\def\vb{{b}}
\def\vc{{c}}
\def\tilpsi{{\tilde\psi}}
\def\tbpsi{{\tilde{\bar\psi}}}

\def\Dslash{{}\hbox{\hskip2pt\vtop
 {\baselineskip23pt\hbox{}\vskip-24pt\hbox{/}}
 \hskip-11.5pt $D$}}
\def\nabslash{{}\hbox{\hskip2pt\vtop
 {\baselineskip23pt\hbox{}\vskip-24pt\hbox{/}}
 \hskip-11.5pt $\nabla$}}
\def\xislash{{}\hbox{\hskip2pt\vtop
 {\baselineskip23pt\hbox{}\vskip-24pt\hbox{/}}
 \hskip-11.5pt $\xi$}}
\def\leftnabla{{\overleftarrow\nabla}}

\def\delL{{\delta_{LL}}}
\def\delG{{\delta_{G}}}
\def\delc{{\delta_{cov}}}

\newcommand {\sqxx}  {\sqrt {x^2+1}}   
\newcommand {\gago}  {\gamma_5}
\newcommand {\Ktil}  {{\tilde K}}
\newcommand {\Ltil}  {{\tilde L}}
\newcommand {\Qtil}  {{\tilde Q}}
\newcommand {\Rtil}  {{\tilde R}}
\newcommand {\Kbar}  {{\bar K}}
\newcommand {\Lbar}  {{\bar L}}
\newcommand {\Qbar}  {{\bar Q}}
\newcommand {\Pp}  {P_+}
\newcommand {\Pm}  {P_-}
\newcommand {\GfMp}  {G^{5M}_+}
\newcommand {\GfMpm}  {G^{5M'}_-}
\newcommand {\GfMm}  {G^{5M}_-}
\newcommand {\Omp}  {\Omega_+}    
\newcommand {\Omm}  {\Omega_-}
\def\Aslash{{}\hbox{\hskip2pt\vtop
 {\baselineskip23pt\hbox{}\vskip-24pt\hbox{/}}
 \hskip-11.5pt $A$}}
\def\Rslash{{}\hbox{\hskip2pt\vtop
 {\baselineskip23pt\hbox{}\vskip-24pt\hbox{/}}
 \hskip-11.5pt $R$}}
\def\kslash{
{}\hbox       {\hskip2pt\vtop
                   {\baselineskip23pt\hbox{}\vskip-24pt\hbox{/}}
               \hskip-8.5pt $k$}
           }    
\def\qslash{
{}\hbox       {\hskip2pt\vtop
                   {\baselineskip23pt\hbox{}\vskip-24pt\hbox{/}}
               \hskip-8.5pt $q$}
           }    
\def\dslash{
{}\hbox       {\hskip2pt\vtop
                   {\baselineskip23pt\hbox{}\vskip-24pt\hbox{/}}
               \hskip-8.5pt $\partial$}
           }    
\def\dbslash{{}\hbox{\hskip2pt\vtop
 {\baselineskip23pt\hbox{}\vskip-24pt\hbox{$\backslash$}}
 \hskip-11.5pt $\partial$}}
\def\Kbslash{{}\hbox{\hskip2pt\vtop
 {\baselineskip23pt\hbox{}\vskip-24pt\hbox{$\backslash$}}
 \hskip-11.5pt $K$}}
\def\Ktilbslash{{}\hbox{\hskip2pt\vtop
 {\baselineskip23pt\hbox{}\vskip-24pt\hbox{$\backslash$}}
 \hskip-11.5pt ${\tilde K}$}}
\def\Ltilbslash{{}\hbox{\hskip2pt\vtop
 {\baselineskip23pt\hbox{}\vskip-24pt\hbox{$\backslash$}}
 \hskip-11.5pt ${\tilde L}$}}
\def\Qtilbslash{{}\hbox{\hskip2pt\vtop
 {\baselineskip23pt\hbox{}\vskip-24pt\hbox{$\backslash$}}
 \hskip-11.5pt ${\tilde Q}$}}
\def\Rtilbslash{{}\hbox{\hskip2pt\vtop
 {\baselineskip23pt\hbox{}\vskip-24pt\hbox{$\backslash$}}
 \hskip-11.5pt ${\tilde R}$}}
\def\Kbarbslash{{}\hbox{\hskip2pt\vtop
 {\baselineskip23pt\hbox{}\vskip-24pt\hbox{$\backslash$}}
 \hskip-11.5pt ${\bar K}$}}
\def\Lbarbslash{{}\hbox{\hskip2pt\vtop
 {\baselineskip23pt\hbox{}\vskip-24pt\hbox{$\backslash$}}
 \hskip-11.5pt ${\bar L}$}}
\def\Rbarbslash{{}\hbox{\hskip2pt\vtop
 {\baselineskip23pt\hbox{}\vskip-24pt\hbox{$\backslash$}}
 \hskip-11.5pt ${\bar R}$}}
\def\Qbarbslash{{}\hbox{\hskip2pt\vtop
 {\baselineskip23pt\hbox{}\vskip-24pt\hbox{$\backslash$}}
 \hskip-11.5pt ${\bar Q}$}}
\def\Acalbslash{{}\hbox{\hskip2pt\vtop
 {\baselineskip23pt\hbox{}\vskip-24pt\hbox{$\backslash$}}
 \hskip-11.5pt ${\cal A}$}}

\begin{flushright}
December 2000\\
hep-th/0012255 \\
US-00-11
\end{flushright}

\vspace{0.5cm}

\begin{center}

{\Large\bf 
Pole Solution in Six Dimensions and\\
Mass Hierarchy}
\footnote{
Based on the talk at the Sendai Phenomenology
Workshop "New Direction to Unified Theories",Oct.23-25,
2000,Aoba Memorial Hall,Tohoku Univ.,Sendai,Japan
}

\vspace{1.5cm}
{\large Shoichi ICHINOSE
         \footnote{
E-mail address:\ ichinose@u-shizuoka-ken.ac.jp
                  }
}
\vspace{1cm}

{\large 
Laboratory of Physics, \\
School of Food and Nutritional Sciences, \\
University of Shizuoka,
Yada 52-1, Shizuoka 422-8526, Japan          }

\end{center}

\vfill

{\large Abstract}\nl
A solution of the 6D gravitational model, 
which has the pole
configuration, is found. The vacuum
setting is done by the 6D Higgs potential and the
solution is for a family of the vacuum and 
boundary parameters.
The boundary condition
is solved by the $1/k^2$-expansion (thin pole
expansion) where $1/k$ is the {\it thickness} of the pole. 
The obtained analytic solution is checked by
the numerical method. 
This is a dimensional reduction model from 6D to 4D
by use of the soliton solution (brane world).
It is regarded as a higher dimensional version of
the Randall-Sundrum 5D model. 
The mass hierarchy is examined. 
Especially the {\it geometrical
see-saw} mass relation is obtained. 
Some physical quantities in 4D world such as the Planck mass,
 the cosmological constant, and fermion masses are focussed.
Comparison with the 5D model is made.  

\vspace{0.5cm}

PACS NO:\ 
04.20.Ex 
04.50.+h 
04.25.-g 
11.10.Kk 
11.25.Mj 
11.27.+d 
12.10.Kt 
12.10.-g 
\nl


\section{Introduction}
The higher dimensional approach is a natural way
to analyze the 4D physics in the geometrical standpoint.
The history traces back to Kaluza-Klein in 1921\cite{Kal21,Klei26}
 where the electro-magnetic force is amalgamated into the
gravitational force in the 5D theory. Stimulated by the recent
development of the string and D-brane theories, a new type
 compactification mechanism was invented by Randall and
Sundrum\cite{RS9905,RS9906}. 
The domain wall configuration in 5D space-time,
which is a kink solution in the extra dimension, 
is exploited.
Before Randall-Sundrum(RS), the supersymmetry is the popular
way to extend the standard model. The D-brane inspired model
has provided us with new possibilities for the extension.
It has some advantages in 
the hierarchy problem and the chiral problem.

A moderate view to the string or D-brane theory is
to regard the formalism as a {\it regularization} of the field
theory. In this standpoint, the extendedness, in the space-time, 
of an object is the essence of the {\it singular-free} behavior.
The non-commutative geometry, which is a recent trend
in the field theory, can be regarded as a systematic way
to introduce the extendedness.
In the ordinary (commutative) geometry,
the soliton configuration can do the same role.
We present a 6D soliton solution, and show that it provides
a new dimensional reduction mechanism.

In the research of the domain world physics, the main theoretical
focuses, at present, are 1) Consistency, 2) Stability and 
3) Localization. 
The first one is the basic thing in all theoretical analysis.
In the analysis of the domain world physics, the delta-function
type special configuration (singular space) is often 
assumed without properly solving the field equation
with some boundary condition. 
This is a main origin of the necessity of the consistency
check. 
In some limitted situation, it can be accepted
and is indeed useful in that the whole physical image can be
easily introduced. 
The approach is, however, not admitted
as the right configuration in the general case. 
It is sometimes discussed that some no-go theorem exists
in some models\cite{KL0001,GL0003,MN0007}. 
Especially the rigorous treatment
is required in the analysis related to the cosmological constant
and the boundary condition\cite{SI00apr,SI0107,SI0008}. 
Recently in order to check
such consistency for a wide-range brane-world models, 
some useful sum rule is presented\cite{GKL0011}. The present
approach treats the configuration in the general setting and
the model is based on the field theory. Therefore it
basically passes the consistency check.
Next item is the stability problem. 
The solution should satisfy the stability condition. 
It is generally rather difficult to analyze because
the solution should be examined from some general configuration
depending on parameters such as the distance between the branes.
One interesting idea is proposed by Goldberger and 
Wise\cite{GW9907b},
where the radion field is utilized. In the present approach,
the stability is guaranteed by the {\it boundary condition}. This is
the popular situation in the soliton physics. The third focus
is whether the ordinary fields 
(matter fields, gauge fields, graviton, etc.) are localized
in some narrow region in the extra space. This is necessary
to be consistent with the fact that we have so far not observed
the extra dimensions (requirement for the dimensional reduction). 
In the present analysis, no real matter fields are taken into
account. Only 6D Higgs scalar is introduced to define the (classical)
vacuum clearly. The localization of other fields, 
in the present analysis, is assumed valid based 
on other references\cite{Oda0006,DS0008} and 
on the analogy to the lattice fermion situation. 

We have some reasons to investigate the higher dimensional
generalization of the RS-model. Firstly some references
suggest the difficulty of making a realistic stable 5D
domain world\cite{KL0001,GL0003}. One of main causes
looks its odd dimensionality. Secondly, in the solitonic
approach, various extended objects 
(wall, string, monopole, ...) appear\cite{CS9604}. 
It is a natural direction to investigate the possibility of
other type configuration than wall. 
Thirdly in the work by
Callan and Harvey\cite{CH85}, they interpret the anomaly
phenomena in 4D world in the higher dimensions.
They exploit the physics of fermion zero modes on
strings in 2n+2 dimensions and domain walls in 2n+1
dimensions. The result is
consistent with the anomaly relations between different
dimensions by Alvarez Gaum\'e and Witten\cite{AW83NP}. 
6D model is necessary, in addition to 5D, to understand
the whole anomaly structure. 
We consider the brane world model has 
great advantages about the anomaly and chiral problems.
Finally, in the present rapid progress of the noncommutative
geometry, the extra space has
even dimensions. We suppose the present approach
has something common with such approach.

The 6D models of the hierarchy problem have already been discussed
in various ways\cite{CN99,INOS0004,GS0004,Nih0005}. Main different points
from those works are 
1) the present approach puts emphasis on
the {\it thickness} of the configuration, 
2) the 6D(bulk) Higgs fields are introduced
to clearly define the (classical) vacua, which are necessary
for the soliton boundary conditions,
3) we treat the warp factors and the scalar (Higgs) fields on the
equal footing, 
4) the boundary condition is systematically 
solved and 
5) supersymmetry is not taken into account.

In Sec.2 we introduce the 6D model and derive a solution
of the 6D classical Einstein equation in Sec.3. 
It is a one parameter
family of soliton solution with the pole configuration. 
The asymptotic forms, in the dimensional reduction, 
of the vacuum parameters are obtained in Sec.4.
In Sec.5, some physical quantities, such as the 4D Planck mass
, the 4D cosmological constant and fermion masses, 
are explained. The {\it see-saw} relation is obtained there.
Some order estimation is also done.
The precise form of the solution is obtained
by analytically solving the boundary condition in Sec.6. 
It is demonstrated that
the vacuum parameters are fixed to be some numbers
by one input (boundary) condition.
The obtained solution is further confirmed by
the numerical method in Sec.7. Finally we conclude in Sec.8.
In app.A we explain, in detail, 
the derivation of the solution treated in Sec.3, . 
Some detail about the numerical analysis of Sec.7 is explained
in App.B.

\section{Model set-up}
We start with the 6D gravitational theory, 
where the metric is Lorentzian, 
with the 6D Higgs potential.
\begin{eqnarray}
S[G_{AB},\Phi]=\int d^6X\sqrt{-G} (-\half M^4\Rhat
- G^{AB}\pl_A\Phi^*\pl_B\Phi-V(\Phi^*,\Phi))\com\nn
V(\Phi^*,\Phi)=\frac{\la}{4}(|\Phi|^2-{v_0}^2)^2+\La\com
\label{model1}
\end{eqnarray}
where $X^A (A=0,1,2,3,4,5)$ is the 6D coordinates and we also use
the notation $(X^A)\equiv (x^\m,\rho,\vp), \m=0,1,2,3.$
$x^\m$'s are regarded as our world coordinates, whereas  
$(X^4,X^5)=(\rho,\vp)$ the extra ones. The extra (spacial) coordinates 
are taken to be polar ones:\ 
$0\leq \rho <\infty,\ 0\leq \vp <2\pi$.
$\rho$ will be regarded as a "radius" and $\vp$ as an "angle" around the 4D pole-world.
See Fig.1.
\begin{figure}
\centerline{\epsfysize=6cm\epsfbox{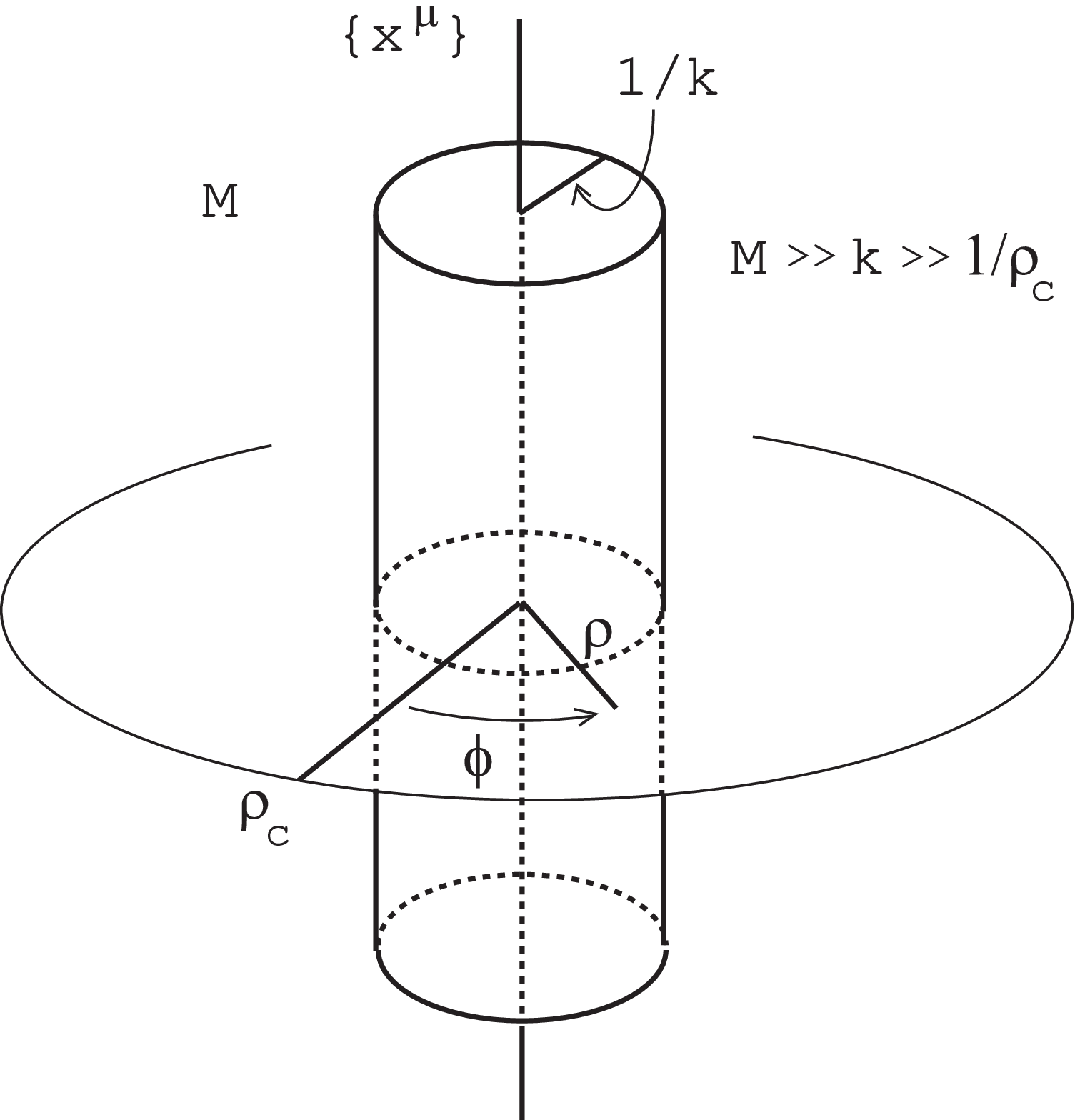}}
   \begin{center}
Fig.1\ The pole configuration.
   \end{center}
\end{figure}
The signature of the 6D metric $G_{AB}$ is $(-+++++)$. 
$\Phi$ is the 6D {\it complex} scalar field, $G=\det G_{AB}$, $\Rhat$ is the
6D Riemannian scalar curvature. $M(>0)$ is the 6D Planck mass
and is regarded as the {\it fundamental scale} of this dimensional reduction
scenario. $V(\Phi^*,\Phi)$ is the Higgs potential and defines
the (classical) vacuum in the 6D world. 
The three parameters $\la,\vz$ and $\La$ in the potential $V$ are called here
{\it vacuum parameters}. 
$\la(>0)$ is a coupling, $v_0(>0)$ 
is the Higgs field vacuum expectation value, and $\La$ is the 6D cosmological
constant. 
It is later shown that the sign of $\La$ must be negative for
the proposed pole vacuum configuration.
We take the line element shown below.
\footnote{
The same type metric was also taken in Ref.\cite{CN99,GS0004}. 
}
\begin{eqnarray}
{ds}^2=\e^{-2\si(\rho)}\eta_\mn dx^\m dx^\n+{d\rho}^2
+\rho^2\e^{-2\om(\rho)}d\vp^2\com\nn
0\leq \rho<\infty\com\q 0\leq\vp<2\pi\com
\label{model2}
\end{eqnarray}
where $\eta_\mn=\mbox{diag}(-1,1,1,1)$. 
This is a natural 6D 
minimal-extension of the original 5D model by Randall-Sundrum
\cite{RS9905}. 
For $\si=\om=0$, $ds^2$ is the 6D Minkowski (flat) space. 
In this choice, the 4D Poincar{\' e}
invariance is preserved. Two "warp" factors $\e^{-2\si(\rho)}$ and
$\e^{-2\om(\rho)}$ appear and play an important
role throughout this paper.
Note that, for the fixed $\rho$ case ($d\rho=0$), the metric is the
{\it Weyl transformation} of the product space of
the flat (Minkowski) space, $\eta_\mn dx^\m dx^\n$, 
and the circle $S^1$, $\rho^2d\vp^2$. 
The two Weyl factors serve for {\it scaling} each space.
The coordinate $\rho$ is regarded as the scaling parameter.

From (\ref{model2}) we can read
\begin{eqnarray}
\left(
\begin{array}{ccc}
\mbox{} & \mbox{} & \mbox{} \\
\mbox{} & G_{AB} & \mbox{} \\
\mbox{} & \mbox{} & \mbox{}
\end{array} 
\right)
=
\left(
\begin{array}{ccc}
\e^{-2\si}\eta_\mn & 0 & 0 \\
0 & 1 & 0 \\
0 & 0 & \rho^2\e^{-2\om}
\end{array} 
\right)
\pr
\label{model2b}
\end{eqnarray}
The 6D Riemannian scalar curvature $\Rhat$ is given by
\begin{eqnarray}
\Rhat=4(-2\si''+5(\si')^2-2\frac{\si'}{\rho})+8\om'\si'
+2(-\om''+(\om')^2-2\frac{\om'}{\rho})
\ ,
\label{model3}
\end{eqnarray}
where $'=\frac{d}{d\rho}$. 
We will later require {\it there are no curvature singularities anywhere}
because the configuration considered has the {\it finite thickness}
(radius of the pole) which should smooth the curvature.
Since $M$-dependence can be absorbed by a simple scaling
($\Phi=M^{2}{\tilde \Phi}, 
v_0=M^{2}{\tilde v_0}, \la=M^{-2}{\tilde \la},
\La=M^{6}{\tilde \La}, x^\m=M^{-1}{\tilde x^\m}, \rho=M^{-1}{\tilde \rho},
\vp={\tilde \vp}$), 
it is sometimes useful to take
\begin{eqnarray}
M=1
\pr\label{model4}
\end{eqnarray}
We will, however, 
explicitly write $M$ as much as possible
in order for the reader not to bother about
the superficial dimensional mismatch.

\section{A solution}
Let us solve the 6D Einstein equation.
\begin{eqnarray}
M^4(\Rhat_{MN}-\half G_{MN}\Rhat )=\nn
-\pl_M\Phi^*\,\pl_N\Phi
-\pl_N\Phi^*\,\pl_M\Phi
+G_{MN}(G^{KL}\pl_K\Phi^*\,\pl_L\Phi+V)\com\nn
\na^2\Phi=\frac{\del V}{\del \Phi^*}\pr
\label{sol1}
\end{eqnarray}
As for the complex scalar field $\Phi$, first of all, it should satisfy
the periodicity condition:\ 
$
\Phi(x,\rho,\vp)=\Phi(x,\rho,\vp+2\pi)
$.
Following Callan and Harvey\cite{CH85},
we consider a simple case that $\Phi$ depends only on 
the extra coordinates 
$\vp$ 
and $\rho$ as
\begin{eqnarray}
\Phi_m(\rho,\vp)= P(\rho)\e^{im\vp}\com\q
m=0,\pm1,\pm2,\cdots\pr
\label{sol3}
\end{eqnarray}
Then the Einstein equation (\ref{sol1}) reduces to
\begin{eqnarray}
M=\m,N=\n\qqqq\qqqq\qqqq\qqqq\qqqq\qqqq\nn
3\si''-6{\si'}^2+3\frac{\si'}{\rho}-3\si'\om'+\om''-(\om')^2
+2\frac{\om'}{\rho}\nn
=M^{-4}({P'}^2+m^2\frac{\e^{2\om}}{\rho^2}P^2+V)\com\label{sol4a}\\
M=\rho,N=\rho\qqqq\qqqq\qqqq\qqqq\qqqq\qqqq\nn
-6{\si'}^2+4\frac{\si'}{\rho}-4\si'\om'
=M^{-4}(-{P'}^2+m^2\frac{\e^{2\om}}{\rho^2}P^2+V)\com\label{sol4b}\\
M=\vp,N=\vp\qqqq\qqqq\qqqq\qqqq\qqqq\qqqq\nn
\mbox{Eq.(C)}\q\q 4\si''-10{\si'}^2
=M^{-4}({P'}^2-m^2\frac{\e^{2\om}}{\rho^2}P^2+V)\com\label{sol4c}\\
\mbox{Matter Eq.}\qqqq\qqqq\qqqq\qqqq\qqqq\qqqq\nn
P''-m^2\frac{\e^{2\om}}{\rho^2}P-(4\si'-\frac{1}{\rho}+\om')P'
=\frac{\la}{2}(P^2-\vz^2)P\ .\label{sol4d}
\end{eqnarray}
The matter equation (\ref{sol4d}) is derived from other three equations.
\footnote{
It reflects the following well-known fact. 
The matter equation $\na^2\Phi=\frac{\del V}{\del \Phi^*}$ in (\ref{sol1})
can also be derived from the (6D) energy-momentum conservation
$M^4\na^M(\Rhat_{MN}-\half G_{MN}\Rhat )=\na^M T_{MN}=0$, where
$T_{MN}=$RHS of the first equation of (\ref{sol1}).
}
From (\ref{sol4a}) and (\ref{sol4b}), and from (\ref{sol4b}) and (\ref{sol4c}),
we obtain two convenient equations as follows.
\begin{eqnarray}
\mbox{Eq.(A)}\q\q 3\si''-\frac{\si'}{\rho}+\si'\om'+\om''-(\om')^2
+2\frac{\om'}{\rho}
=2M^{-4}{P'}^2\com\label{sol5a}\\
\mbox{Eq.(B)}\q\q -16{\si'}^2+4\frac{\si'}{\rho}-4\si'\om'+4\si''
=2M^{-4}V\pr\label{sol5b}
\end{eqnarray}
The three equations Eq.(A),Eq.(B) and Eq.(C)
may be considered as those to be solved. 
They have three unknown functions $P(\rho), \si(\rho), \om(\rho)$ and
one unknown integer parameter $m$. 
We call this set of the three
equations {\it reduced Einstein equations} (REE).

Now we must consider the boundary condition(asymptotic behavior), 
at $\rho=\infty$(infrared region), 
for $P(\rho),\si(\rho)$ and $\om(\rho)$. As for $P(\rho)$, we naturally
take
\begin{eqnarray}
\rho\ra\infty\com\q P(\rho)\ra\pm\vz
\pr\label{sol6}
\end{eqnarray}
The plural sign $\pm$ comes from the fact that the Higgs potential $V(\Phi)$ is 
the {\it even} function of $\Phi$. 
Later soon, we will choose the plus sign above by use of
the freedom of an arbitrary parameter $e_0$( defined later ). 
As for other two, we assume (from the "experience" in 5D Randall-Sundrum
model\cite{RS9905,RS9906,SI00apr}) 
$\si'\ra a(\mbox{const}), \om'\ra b(\mbox{const})$ as $\rho\ra\infty$. 
Then from the REE,
we can deduce
\begin{eqnarray}
m=0\com\nn
\mbox{As} \q\rho\ra\infty\com\q
\si'\ra {\tilde \al}\com\q \om'\ra {\tilde \al}\com\q
{\tilde \al}=\pm\sqrt{\frac{-\La}{10}}M^{-2}
\pr\label{sol7}
\end{eqnarray}
As for the plural sign above, we take the plus one
which damps the scaling behavior (\ref{model2}) in the
asymptotic region. 
Note that the present asymptotic requirement demands
the {\it isotropic property} around the $\rho=0$ axis
($m=0$), that is, $\Phi$ should not depend on the direction $\vp$,  
that is, the {\it pole} configuration. See Fig.1.
We take the {\it plus sign} above in order for this mechanism
to work, in the infrared region, 
for {\it exponentially suppressing} the 4D part of the metric
(\ref{model2}). 
In the above result we must have the condition:
\begin{eqnarray}
\La<0
\pr\label{sol7b}
\end{eqnarray}
The geometry should be Anti de Sitter in the (infrared)
asymptotic region. We should further note that, in the infrared
region above, the 6D scalar curvature is a positive constant:\ 
${\hat R}\ra 30{\tilde \al}^2=-3\La M^{-4}>0$. 

Next we must consider the boundary condition  
at $\rho=+0$(ultra-violet region). Assuming the following things:\ 
1)\ As $\rho\ra +0$,
the three unknown functions $\si',\om'$ and $P$ goes like
\ $\si'\ra s\rho^a\ ,\ \om'\ra w\rho^b\ ,\ P\ra x\rho^c$ where
$s,w$ and $x$ are {\it non-zero} constants ($s\neq 0,w\neq 0,x\neq 0$);\ 
2)\ The 6D scalar curvature $\Rhat$ is {\it regular everywhere};\ 
3)\footnote
{
The assumption $a=b$ is based on the expected "duality"
between IR and UV behaviors. Note that, in the IR, $\si'$ and $\om'$
behave similarly in (\ref{sol7}).
}
\ $a=b$,
we can deduce, from REE, that there are two cases:\  
\begin{eqnarray}
\mbox{Case 1}\q
a=b=1\ ,\ c=0\ ,\ s=\frac{\la}{16}(x^2-\vz^2)^2+\frac{\La}{4}\label{sol8a}\\
\mbox{Case 2}\q
a=b=1\ ,\ c>1\ ,\ s=\frac{\la}{16}\vz^4+\frac{\La}{4}\label{sol8b}
\end{eqnarray}
If we put off the condition $x\neq 0$ in (\ref{sol8a}) and take $x=0$,
eq.(\ref{sol8a}) reduces to (\ref{sol8b}). 

Let us take the following form for $\si'(\rho),\om'(\rho)$ and $P(\rho)$
as a solution
\footnote{
Some properies of the form of the series (\ref{sol9})
in relation to the important equation (\ref{app3}) is
described in App.A.
}
.
\begin{eqnarray}
\si'(\rho)=\al\sum_{n=0}^\infty\frac{c_{2n+1}}{(2n+1)!}\{\tanh (k\rho)\}^{2n+1}\com\nn
\om'(\rho)=\al\sum_{n=0}^\infty\frac{d_{2n+1}}{(2n+1)!}\{\tanh (k\rho)\}^{2n+1}\com\nn
P(\rho)=v_0\sum_{n=0}^\infty\frac{e_{2n}}{(2n)!}\{\tanh (k\rho)\}^{2n}\com
\label{sol9}
\end{eqnarray}
where 
$\al\equiv +\sqrt{\frac{-\La}{10}}M^{-2}$ 
and 
$c$'s, $d$'s and $e$'s are coefficient-constants 
(independent of $\rho$) to be determined. 
Because REE have the "formal" symmetry of {\it parity}:\ $\rho\ra -\rho$, 
we know the solutions are parity odd or even functions.
(Note that $0\leq \rho <\infty$, (\ref{model2}). )
In (\ref{sol9}), $\si'$ and $\om'$ are composed of {\it odd} powers of
tanh($k\rho$), whereas $P$ is of {\it even} powers. 
The choice is deduced by the behavior at the ultra-violet region
(\ref{sol8a},\ref{sol8b}).
The expansion (\ref{sol9}) could be regarded as a sort of "Kaluza-Klein mode expansion".
\footnote{
In comparison with the work by Goldberger and Wise\cite{GW9907a}, 
not only
the scalar field $P$ but also the warp factors $\si'$ and $\om'$ are
"Kaluza-Klein expanded".
}
In the present case, however, there is {\it no periodicity} in the $\rho$-coordinate.
We {\it neither} have the {\it translation invariance} 
$\rho\ra\rho+\mbox{const.}$, which should
be compared with 5D case\cite{SI00apr,SI0107}. 
A {\it new mass scale} $k(>0)$ is introduced here to make 
the quantity  $k \rho$ dimensionless. The physical meaning of $1/k$ 
is the "thickness" of the pole. 
The parameter $k$, with $M$ and $\rho_c$(defined later), plays a central role
in this dimensional reduction scenario.
We call them, $(k,M,\rho_c)$, {\it fundamental parameters}.  
The distortion of 6D space-time by the
existence of the pole should be sufficiently small so that 
the quantum effect of 6D gravity
can be ignored and the present {\it classical} analysis is valid. This requires
the condition\cite{RS9905}
\begin{eqnarray}
k\ll M\pr
\label{sol10}
\end{eqnarray}

The boundary conditions (\ref{sol6},\ref{sol7}) with the $+$ choice in the plural signs, require
the coefficient-constants to have the following constraints 
\begin{eqnarray}
1=\sum_{n=0}^\infty\frac{c_{2n+1}}{(2n+1)!}\com\q
1=\sum_{n=0}^\infty\frac{d_{2n+1}}{(2n+1)!}\com\q
1=\sum_{n=0}^\infty\frac{e_{2n}}{(2n)!}\com
\label{sol11}
\end{eqnarray}
, which are obtained by considering
the asymptotic behaviors
$\rho\ra \infty$ in (\ref{sol9}). We will use these constraints
in Sec.6.
In App.A, we explain the determination of all coefficients
in (\ref{sol9}) from the REE. 
All ones are fixed except one {\it free} parameter $e_0$($=P(0)/v_0$).
\footnote{
The corresponding {\it global} symmetry looks some 'deformation'
of the constant transformation: $P(\rho)\ra P(\rho)+e_0$.
The symmetry is valid for the special case:\ $m=0, \la=0$. 
This situation should be compared with the 5D case where
the constant translation symmetry for a coordinate exists\cite{SI00apr,SI0107}.
} 
The first two orders are
concretely given as 
\begin{eqnarray}
\left\{
\begin{array}{c}
c_1=\frac{M^{-4}}{4k\al}\{ \fourth\la\vz^4(1-\ez^2)^2+\La \}\com\\
d_1=-\frac{2}{3}c_1\com\\
e_0\ :\ \mbox{free parameter},
\end{array}\right.\nn
\left\{
\begin{array}{c}
c_3=\frac{3}{32}\frac{\la^2\vz^6}{k^3\al}M^{-4}\ez^2(1-\ez^2)^2
+c_1(2+5\frac{\al}{k}c_1)\ ,\\
d_3=-\frac{4}{3}c_1(1+5\frac{\al}{k}c_1)\com \\
e_2=-\fourth\frac{\la\vz^2}{k^2}\ez(1-\ez^2)\com
\end{array}\right.\nn
\label{sol12}
\end{eqnarray}
Here we note that those parts of
the UV boundary condition (\ref{sol8a},\ref{sol8b}), which are not
taken into account in the starting assumption (\ref{sol9}),
are satisfied by the above solution:
\ [Case 1]\ $e_0\neq 0, x=\vz\ez, s=\al c_1;\ 
[Case 2]\ \ez=0, s=\al c_1$. 
The third order is given by
\begin{eqnarray}
e_4\times M^{-4}\{\frac{k\vz^2}{\al}e_2
+\frac{1}{12}\frac{\la\vz^4}{k\al}\ez(1-\ez^2)\}\nn
=(-\frac{4}{9}c_3+2c_1)
+\frac{\al}{k}\cdot\frac{5}{3}(d_1c_3+d_3c_1+2c_1c_3)\nn
-\frac{\la\vz^4}{k\al}\fourth M^{-4}(1-3\ez^2){e_2}^2
+6\frac{k\vz^2}{\al} M^{-4}{e_2}^2\com\nn
c_5=-\frac{5}{12}\frac{\la\vz^4}{k\al}M^{-4}\ez(1-\ez^2)e_4
+\frac{100}{9}c_3+4c_1        \nn
+\frac{\al}{k}\cdot\frac{10}{3}(d_1c_3+d_3c_1+8c_1c_3)
-\frac{5}{4}\frac{\la\vz^4}{k\al}M^{-4}(1-3\ez^2){e_2}^2\com\nn
d_5=-2c_5+\frac{120}{7}\{
\frac{2}{3}\frac{k\vz^2}{\al}M^{-4}e_2e_4+\frac{13}{9}c_3
-\frac{1}{5}c_1+\frac{11}{18}d_3+\frac{2}{5}d_1\nn
-\frac{\al}{k}\cdot\frac{1}{6}(d_1c_3+d_3c_1-2d_1d_3)
-4\frac{k\vz^2}{\al}M^{-4}{e_2}^2
                              \}
\label{sol13}
\end{eqnarray}
Note that, in the first equation, $e_4$ is written by lower-order terms\nl
$(e_0,e_2;c_1,c_3;d_1,d_3)$. Putting this result of $e_4$ into the second
equation, we see $c_5$ is written by the lower-order terms. Similarly
to the third equation with respect to $d_5$. 

The general terms $(c_{2n+1},d_{2n+1},e_{2n}), n\geq 3$ are
obtained in App.A. They are expressed by
the coupled {\it linear} equations
(\ref{app11}) with respect to them
and can be solved recursively in the increasing order
of $n$. 
All coefficients are expressed
by four parameters $\la, \vz, \La$ and $e_0$. (We may take
$M=1$ as explained before. We may also take $k=1$, 
keeping the relation (\ref{sol10}) in mind. This is because
the thickness parameter $k$ can be absorbed by the
change : 
$c_{2n+1}\ra \frac{\al}{k}c_{2n+1}, d_{2n+1}\ra \frac{\al}{k}d_{2n+1},
\La\ra \frac{\La}{k^2}M^{-4}, \la\ra \frac{\la}{k^2}M^4$.
We will, however, keep $k$ as much as possible
in order for the reader not to bother about the
superficial dimensional mismatch.) 
In the above results, we see $e_{2n}$($n\geq 1$) is the odd function of $e_0$, 
whereas $c_{2n+1}$ and $d_{2n+1}$ are the even functions. 
Under the sign change $e_0\ra -e_0$, $P$ changes the sign,
whereas $\si'$ and $\om'$ do not. As commented before, 
we can choose the plus sign
in (\ref{sol6}) by use of this freedom.
The above four parameters have three constraints (\ref{sol11})
from the boundary condition at the infrared infinity. Hence
the present solution is {\it one-parameter family} solution. 
We will solve the constraints in Sec.6. 
As a final note in this section, we point out that, 
in comparison with 5D RS-model\cite{SI00apr,SI0107}, 
there appears no lower-bound for $\La$. 

\section{Meaning of the Solution}
Let us examine the obtained solution in relation to the
two general theorems: Goldstone's theorem and the Derrick's
theorem.\nl
\nl
(i) Spontaneous Symmetry Breakdown\nl
The present model (\ref{model1}) has the discrete symmetry:
\begin{eqnarray}
\Phi\q\ra\q -\Phi\com\q
\Phi^*\q\ra\q -\Phi^*\com
\label{mean1}
\end{eqnarray}
and the continuous (global) symmetry:
\begin{eqnarray}
\Phi\q\ra\q \e^{i\th}\Phi\com\q
\Phi^*\q\ra\q \e^{-i\th}\Phi^*\com
\label{mean2}
\end{eqnarray}
where $\th$ is a constant parameter. (\ref{mean1}) is a special
case of (\ref{mean2}) and appears also in the case of the real
scalar field. If the vacuum breaks the continuous symmetry,
a massless scalar boson 
appears (Goldstone's theorem\cite{GSW62,Cole85}). In the present
case, however, we confine the complex scalar field $\Phi$ to
the real field by taking only the $m=0$ mode in (\ref{sol3}).
(In fact we can start from the real scalar $\Phi$ in (\ref{model1})
and obtain the same result.)
This is required from the demand that the IR boundary condition
of the warp factors is the same as the 5D RS-model. 
Hence we should understand that, in the present solution,
the discrete symmetry (\ref{mean1}) is spontaneously broken,
not the continuous one (\ref{mean2}). No Goldstone boson appears,
and the situation is similar to the the domain wall case of 5D model
\cite{SI00apr,SI0107}. When we want to consider the breaking of
the continuous symmetry, the standard way 
is to take the Abelian Higgs model\cite{GMS0104} where
the Goldstone boson is 'eaten'
\footnote{
The massless Goldstone mode is replaced by one mode of the
massive gauge field.
}
 by the gauge field
(Higgs phenomenon). 
Other approach
is also examined in \cite{CK99PLB}. \nl
\nl
(ii)Role of the Warp Factors\nl
After taking the 'warped' metric (\ref{model2}) and assuming the
form of the complex scalar field (\ref{sol3}) with $m=0$, the action
reduces to
\begin{eqnarray}
S=2\pi\int d^4x\ d\rho\ \rho\e^{-4\si-\om}
\{ 
\e^{2\si}{\dot \Phi}^*{\dot \Phi} -{P'}^2
+2(2\si''-5{\si'}^2+2\frac{\si'}{\rho})\nn
-4\om'\si'
+(\om''-{\om'}^2+2\frac{\om'}{\rho})
-\frac{\la}{4}(P^2-\vz^2)^2-\La
\}
\label{mean3}
\end{eqnarray}
where ${\dot \Phi}=\frac{d\Phi}{dt}$ which vanishes in the present static
solution. If we omit the kinetic term of ${\dot \Phi}^*{\dot \Phi}$
in the above integral formula, 
the remaing part is regarded as (-1)$\times$(Energy of the static solution).
Its finiteness is apparent from the factor of $\e^{-4\si-\om}$ and
the regular behavior of $P, \si'$ and $\om'$ near $\rho=0$.
The expression (\ref{mean3}) looks like an interacting theory of scalars. As for the localized
solutions in scalar theories, there exists the well-known theorem called
Derrick's theorem\cite{Derri64,Cole85}. It says, in the case of the space
dimensions higher than or equal to 2, the only non-singular time-independent
solutions of finite energy are the ground states. The present case escapes
this trivial situation in the following way. 
 As shown in (\ref{mean3}), the interraction forms of the fields $\si(\rho),\om(\rho)$
are not those of scalars but those of dilatons. Especially the linear terms
($\si'', \si'/\rho, \om'', \om'/\rho$) appear. They break the positivity
property used in the proof of the Derrick's theorem.

\section{Vacuum parameters:\ $M$ and $k$-dependence in the
dimensional reduction}
Let us examine the behavior of the vacuum parameters
($\La,\vz,\la$) near the 4D world limit (thin pole limit): 
$k\ra \infty$(the dimensional reduction).
This should be taken consistently with (\ref{sol10}). We will specify
the above limit in the more well-defined way later.
In (\ref{sol9}), we note that $\{\tanh (k\rho)\}^{2n+1}\ra \th(k\rho)\ ,\ 
\{\tanh (k\rho)\}^{2n}\ra \th(k\rho)\ ,\rho\geq 0$ as $k\ra\infty$ .  
Assuming the convergence of the infinite series, we can conclude
that $c_1\sim O(k^0)\ ,\ e_2\sim O(k^0)$ as $k\ra \infty$. 
($e_0$ is a free parameter.)
From the former relation, using the explicit result of $c_1$ (\ref{sol12}), 
we know 
$-\La\sim\la\vz^4\sim M^4k\al$. 
Because $\al\sim\sqrt{-\La}M^{-2}$, this says 
$\sqrt{-\La}\sim kM^2$ and $\la\vz^4\sim k^2M^4$. 
From the latter one, using the explicit result of $e_2$ (\ref{sol12}), 
we know $\la\vz^2\sim k^2$. Hence we obtain
\begin{eqnarray}
-\La\sim M^4k^2\com\q
\vz\sim M^{2}\com\q
\la\sim M^{-4}k^2\q
\mbox{as}\q k\ra \infty\pr
\label{asy1}
\end{eqnarray}
These are the {\it leading} behavior of the vacuum parameters
in the dimensional reduction from 6D to 4D as $k\ra\infty$. 
As for the $k$-dependence, the above result is the {\it same} as
the 5D model of Randall and Sundrum\cite{RS9905,SI00apr}.
The more precise forms of (\ref{asy1}) will
be obtained, in Sec.6, using the constraints (\ref{sol11}).

\section{Physical constants and See-Saw relation}
In order to express some physical scales in terms of
the fundamental parameters $M$, $k$ and $\rho_c$(
to be introduced soon), we consider the case that
the 4D geometry is slightly fluctuating around 
the Minkowski (flat) space.
\begin{eqnarray}
{ds}^2=\e^{-2\si(\rho)}g_\mn dx^\m dx^\n+{d\rho}^2
+\rho^2\e^{-2\om(\rho)}d\vp^2\com\nn
g_\mn=\eta_\mn+h_\mn\com\q h_\mn\sim O(\frac{1}{k})
\pr
\label{para1}
\end{eqnarray}
The leading order $O(k^0)$ results of the previous section remain
valid.

\subsection{The Planck mass}
The gravitational part of 6D action (\ref{model1}) reduces to
4D action as
\begin{eqnarray}
\int d^6X\sqrt{-G}M^4\Rhat\sim
M^4\int_0^{2\pi}d\vp\int_{0}^{\rho_c}d\rho\,\rho\e^{-2\si(\rho)-\om(\rho)}
\int d^4x\sqrt{-g}R+\cdots\com
\label{para2}
\end{eqnarray}
where the {\it infrared regularization} parameter $\rho_c$
is introduced. $\rho_c$ specifies the {\it size} of the extra 2D space.
Using the asymptotic forms, $\si\sim \al \rho,\ \om\sim \al \rho$ 
as $\rho\ra \infty$ and
$\al=
\sqrt{\frac{-\La}{10}}M^{-2}\sim k$ as $k\ra\infty$, 
we can evaluate the order of $M_{pl}$ as
\footnote{
A similar relation is derived in other 6D model\cite{GS0004}
}
\begin{eqnarray}
{M_{pl}}^2\sim M^42\pi\int_{0}^{\rho_c}d\rho\,\rho\,\e^{-3\al\rho}
\sim \frac{M^4}{\al^2}\sim\frac{M^4}{k^2}\com\label{para3}
\end{eqnarray}
where we have used the {\it 4D reduction condition}:
\begin{eqnarray}
k\rho_c\gg 1\pr\label{para3b}
\end{eqnarray}
The result (\ref{para3}) is {\it different} from
5D model of Randall-Sunsrum \cite{RS9905,SI00apr}:\ 
${M_{pl}}^2\sim\frac{M^3}{k}$.

The above condition (\ref{para3b})
should be interpreted as the precise (well regularized) definition
of $k\ra\infty$ (dimensional reduction) used so far. 
We note $\rho_c$ dependence in (\ref{para3})
is negligible for $k\rho_c\gg 1$. This behavior shows the distinguished
contrast with the Kaluza-Klein reduction of 6D to 4D
(${M_{pl}}^2\sim M^4(\rho_c)^2$)
as stressed in \cite{RS9905} for the 5D case.

Writing (\ref{para3}) as
\begin{eqnarray}
\frac{M_{pl}}{M}\sim\frac{M}{k}\com
\label{para3c}
\end{eqnarray}
we notice this mass relation is the {\it geometrical see-saw}
relation corresponding to the matrix\ :
\begin{eqnarray}
\left(
\begin{array}{cc}
0 & M \\
M & M_{pl}
\end{array}
\right)
\pr
\label{para3d}
\end{eqnarray}
This provides the geometrical approach to the see-saw mechanism
\cite{Yana79,GRS79}
which is usually explained by the diagonalization of the (neutrino)
mass matrix. ( See a textbook\cite{MP91}.)
For some discussion about the neutrino mass, using the above relation,
see the final section.

\subsection{The cosmological term}
The cosmological part of (\ref{model1}) reduces to 4D action as
\begin{eqnarray}
\int d^6X\sqrt{-G}\La\sim
\La\int_0^{2\pi}d\vp\int_{0}^{\rho_c}d\rho\,
\rho\,\e^{-4\si(\rho)-\om(\rho)}\int d^4x\sqrt{-g}
\equiv \La_{4d}\int d^4x\sqrt{-g}\com\nn
\La_{4d}\sim 
\frac{\La}{\al^2}\sim -M^4<0\com\q
kr_c\gg 1\pr\label{para4}
\end{eqnarray}
$\La_{4d}$ is the cosmological term in the 4D space-time. 
It does not, like $M_{pl}$, depend on $\rho_c$ strongly.
The above result differs from RS-model in that 
$k$-dependence disappears. 
(cf. $\La_{4d}\sim -M^3k$ in RS-model\cite{SI00apr}.) 
The result says the 4D space-time should also be {\it anti de Sitter}.

\subsection{Numerical fitting of $M, k$ and $\rho_c$}
Let us examine what orders of values should we take for the fundamental
parameters $M$ and $k$. 
( $\rho_c$ is later fixed by the information of the 4D fermion masses. )
Using the value $M_{pl}\sim 10^{19}$GeV , the "rescaled" cosmological
parameter ${\tilde \La}_{4d}\equiv \La_{4d}/{M_{pl}}^2$ \cite{foot5a}
has the relation:
\begin{eqnarray}
\sqrt{-{\tilde \La}_{4d}}\sim k\sim M^2\times {10}^{-19}\ \mbox{GeV}\com
\label{para5}
\end{eqnarray}
where the relations (\ref{para3}) and (\ref{para4}) are used. 
The unit of mass is GeV and
this unit is taken in the following. 
The observed value of ${\tilde \La}_{4d}$
is not definite, even for its sign.\cite{Perl99} 
If we take into account the quantum effect, the value of
${\tilde \La}_{4d}$ could run along the renormalization\cite{foot5b}.
Furthermore if we consider
the parameter ${\tilde \La}_{4d}$ represents some "effective"
value averaging over all matter fields, 
the value, no doubt, changes during
the evolution of the universe. (Note the model (\ref{model1})
has no (ordinary) matter fields.) 
Therefore, instead of specifying 
${\tilde \La}_{4d}$, it is useful to consider various possible cases
of ${\tilde \La}_{4d}\sim -k^2$.

Some typical cases are
1) ($k={10}^{-41}, M={10}^{-11}$),\ 
2) ($k={10}^{-13}, M={10}^3$)\ 
3) ($k=10, M={10}^{10}$)\ 
4) ($k={10}^4, M={10}^{11.5}$)\ 
and 
5) ($k={10}^{19}, M={10}^{19}$).
Case 1) gives the most plausible present value of the cosmological constant.
The wall thickness
$1/k={10}^{41}$[GeV$^{-1}$]
, however, is the radius of the present universe.
This implies the extra dimensional effect appears
at the cosmological scale, which should be abandoned 
in the usual sense. 
For further discussion, see the final section. 
Case 2) gives  $1/k={10}^{13}$ GeV$^{-1}$ $\sim 1$mm which is
the minimum length at which the Newton's law is checked\cite{ADD98,AADD98}. 
Usually $k$ should be larger than this value so that
we keep the observed Newton's law.
5) is an extreme case $M=M_{pl}$. 
The fundamental scale $M$ is given by the Planck mass. 
In this case, $\rho_c\gg 1/k=1/M_{pl}$
is acceptable, whereas $\sqrt{-{\tilde \La}_{4d}}\sim M_{pl}$
is completely inconsistent with the experiment and
requires explanation. 
Most crucially the condition (\ref{sol10}) breaks down.
Cases 3) and 4) are some intermediate cases which are acceptable except
for the cosmological constant. They will be used in the next
paragraph. 
At present any choice of ($k,M$)
looks to have some trouble if we take into account the cosmological
constant. We consider the observed cosmological constant
($10^{-41}$GeV) should be explained by some unknown mechanism. 

As in the Callan and Harvey's paper\cite{CH85}, we can
have the 4D {\it massless chiral} fermion bound to the wall
by introducing 6D {\it Dirac} fermion $\psi$ into (\ref{model1}).
\begin{eqnarray}
S[G_{AB},\Phi]+\int d^6X\sqrt{-G}(\psibar\nabslash\psi
+g\psibar (\mbox{Re}\Phi-i\Ga^7\mbox{Im}\Phi) \psi)
\pr
\label{conc1}
\end{eqnarray}
If we {\it regulate} the extra axis by the finite range $0\leq y\leq \rho_c$,
the 4D fermion is expected to have a small mass 
$m_f\sim k\e^{-k\rho_c}$ 
(This phenomenon is known in the condensed matter physics
as the surface mode. It 
is also confirmed in the two-walls configuration of the 5D lattice theory
\cite{Sha93,Vra98}).
If we take the case 3) in Subsec.5.3 ($k=10,M={10}^{10}$)
and regard the 4D fermion as a neutrino ($m_\nu\sim {10}^{-11}-{10}^{-9}\mbox{GeV}$),
we obtain $\rho_c=2.76-2.30 \mbox{GeV}^{-1}$. 
If we take case 4) ($k={10}^4,M={10}^{11.5}$), we obtain
$\rho_c=(3.45-2.99)\times {10}^{-3}\mbox{GeV}^{-1}$. 
When the quarks or other leptons ($m_q,m_l\sim 10^{-3}-10^2\mbox{GeV}$) are taken
as the 4D fermion, and take the case 4) in Subsec.5.3,
we obtain $\rho_c=(1.61-0.461)\times {10}^{-3}$GeV$^{-1}$.
It is quite a fascinating idea
to {\it identify the chiral fermion zero mode bound to the pole
with the neutrinos, quarks or other leptons}.

\subsection{The valid region of the present approach}
The condition on $k$ in the present model, from
(\ref{sol10}) and (\ref{para3b}), is given as
\begin{eqnarray}
\frac{1}{\rho_c}\ll k \ll M\pr
\label{chi2}
\end{eqnarray}
In ref.\cite{SI00apr}, it is pointed out that 
the present mechanism has the similarity to the 
chiral fermion determinant. The above relation corresponds to
$|k^\m|\ll M_F \ll \frac{1}{t}$
in ref.\cite{SI98,SI99}.
Especially the thickness parameter $k$ corresponds to
5D fermion mass parameter $M_F$ which is
a very important parameter, in the lattice simulation, 
to give good numerical results 
for the soft-hadron physics (such as the pion mass).

\section{Precise form of Solution (Analytical Approach)}
Let us determine the precise form of the solution
by solving the boundary condition (\ref{sol11}). 
An interesting aspect of the present
 solution is that some family of vacua is selected as the
consistent (classical) configuration. 
In other words, the vacuum parameters are fixed to be
some numbers (if one free parameter is fixed). 
Let us determine the precise
form of (\ref{asy1}) and (\ref{sol9}) 
using the three constraints (\ref{sol11}).
The coefficients ($c_{2n+1},d_{2n+1},e_{2n}$) of (\ref{sol9})
are solved, in (\ref{sol12}) and (\ref{sol13}), 
in terms of four parameters $\la,\vz,\La$ and $e_0$. 
Therefore the present solution is {\it one parameter}
family solution. 
The precise
forms are obtained by the $\frac{1}{k^2}$-expansion
for the case $k\rhc\gg 1$ as
\begin{eqnarray}
\sqrt{-\La}=M^2k(\al_0+\frac{\al_1}{(k\rhc)^2}+\cdots)
=M^2 k\sum_{n=0}^{\infty}\frac{\al_n}{(k\rhc)^{2n}}\com\q\nn
\la\vz^4=M^4k^2(\del_0+\frac{\del_1}{(k\rhc)^2}+\cdots)
=M^4k^2\sum_{n=0}^{\infty}\frac{\del_n}{(k\rhc)^{2n}}\com\q\nn
\vz^2=M^{4}(\be_0+\frac{\be_1}{(k\rhc)^2}+\cdots)
=M^{4}\sum_{n=0}^{\infty}\frac{\be_n}{(k\rhc)^{2n}}
\com
\label{vac1}
\end{eqnarray}
where $\al$'s,$\del$'s and $\be$'s are some numerical (real) numbers
to be consistently chosen using (\ref{sol11}).

If we assume the infinite series
of (\ref{sol11}) converge  sufficiently rapidly,we 
can safely truncate it at the first few terms. 
In order to demonstrate how the vacuum parameters are fixed, 
we take into account up to $n=2$ in (\ref{sol11}) and 
the leading order in (\ref{vac1}).
\footnote{
In ref.\cite{SI0107}, the improved calculation results are
obtained. It takes into account terms up to n=6th order.
}
  We present two
sample solutions:\ 
\begin{eqnarray}
\mbox{Solution 1  }  e_0=-0.5\mbox{(input)}, \nn
(\al_0,\be_0,\del_0)=
(2.507019, 1.113947, 29.14319)\ ;\nn 
\mbox{Solution 2  }  e_0=-0.8\mbox{(input)}, \nn
(\al_0,\be_0,\del_0)=
(1.864786, 0.6424929, 15.90691). \nn
\label{vac2}
\end{eqnarray}
For each solution, the expansion
coefficients
\footnote{
When we see the convergence behavior of the series
appearing in (\ref{sol9}) and (\ref{sol11}),
we should take into account the factorial factors
which divide the all coefficients c's, d's and e's. 
}
 are obtained as
\begin{eqnarray}
\mbox{Solution 1  } \qqqq\qqqq\qqqq\qqqq\nn
(c_1,c_3,c_5)=(-0.689617,13.1849,-60.9444),\nn
 (d_1,d_3,d_5)=(0.459745,-1.59403,96.7112),\nn
 (e_0,e_2,e_4)=(-0.5,2.45270,6.56764);\nn \nn
\mbox{Solution 2  } \qqqq\qqqq\qqqq\qqqq\nn
(c_1,c_3,c_5)=(-1.25575,7.33112,124.067),\nn
 (d_1,d_3,d_5)=(0.837165,-4.52497,110.039),\nn
 (e_0,e_2,e_4)=(-0.8,1.78258,21.809). \nn
\label{vac3}
\end{eqnarray}
For the solution 2, we plot $P(\rho), \si'(\rho), \om'(\rho)$
and $\Rhat(\rho)$ in Fig.2, Fig.3, Fig.4 and Fig.5, respectively.
We stress that the Riemann scalar curvature $\Rhat$ is
{\it everywhere non-singular}. Note that
$\Rhat/\al^2\ra -12\sqrt{10}c_1/A,\ A\equiv \sqrt{-\La}/(M^2k)$
as $\rho\ra +0$ (ultraviolet region). (Purely for the technical
reason, Fig.5 do not show the value of $\Rhat$ at $\rho=0$.)
As for the infrared region, we know
$\Rhat/\al^2\ra 30$ as $\rho\ra \infty$.

Note the following points:\ 
1) The above results are valid for general fundamental
parameters $(k,M,\rho_c)$ except the thin-pole condition
$k\rho_c\gg 1$;\ 
2) For each given value of $e_0$, the solution looks unique
(when a solution exists) 
as far as the reasonable range of $(\al_0,\be_0,\del_0)$
is concerned;\ 
3) For the choice of the plus boundary for the Higgs field $P$ 
(the plus sign in (\ref{sol6})), we do not find solutions
for positive $e_0$ in the reasonable range of parameters;\ 
4) For one given input value ( $e_0$ in the above case. The 
6D Higgs mass (in unit $k$) 
$\sqrt{\la\vz^2}/k\sim \sqrt{\del_0/\be_0}$ is another
example of such value.)
the cosmological constant (in unit $M^2k$) 
$\sqrt{-\La}/M^2k\sim\al_0$ is fixed
\footnote{
In some references, this is mistakenly called "fine-tuning".
The cosmological constant is here not fine-tuned but fixed
from the (classical) field equation and the input value.
}
;\ 
5) The above very fine results are obtained by the standard
method of the numerical calculus:\ {\it Newton method}. 

Our solution
has one {\it free parameter}.
Using this freedom we can adjust one of the three vacuum parameters
in the way the observed physical values are explained.
In (\ref{vac2}), we take $e_0=P(0)/\vz$ as the adjustment.
Solution 2 has the vacuum expectation value
$v_0M^{-2}\sim\sqrt{\be_0}=0.802$, the cosmological constant
$\La M^{-4}k^{-2}\sim -{\al_0}^2=-3.48$ and 
the 6D Higgs mass $\sqrt{\la\vz^2}k^{-1}\sim \sqrt{\del_0/\be_0}=4.98$.
We notice the dimensionless vacuum parameters, 
$v_0M^{-2},\ \La k^{-2}M^{-4}$ and $\la k^{-2}M^4$, 
are specified only by the value of $k\rho_c$ and one input data, say, $e_0$.
When we further specify $k$ and $M$, as considered in Subsec 5.3, 
the values $v_0,\ \La$ and $\la$ are themselves obtained.
Any higher-order, in principle, can be obtained by
the $\frac{1}{k^2}$-expansion.
 
\begin{figure}
\centerline{\epsfysize=4cm\epsfbox{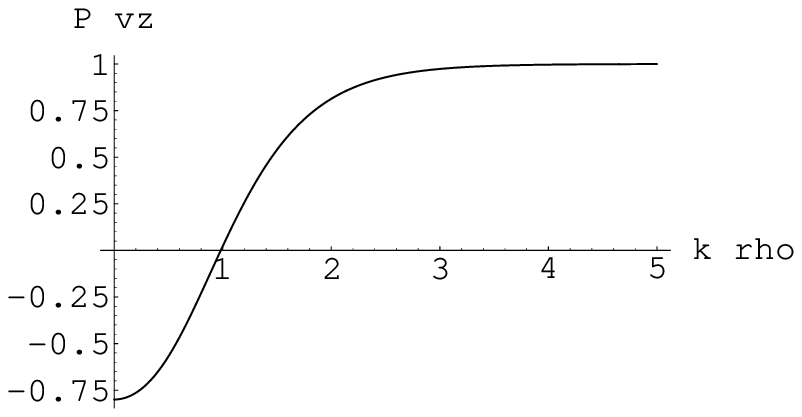}}
   \begin{center}
Fig.2\ The Higgs Field $P(\rho)/\vz$,
(\ref{sol9}). Solution 2 in (\ref{vac2}) and (\ref{vac3}). 
Horizontal axis: $k\rho$.
   \end{center}
\end{figure}

\begin{figure}
\centerline{\epsfysize=4cm\epsfbox{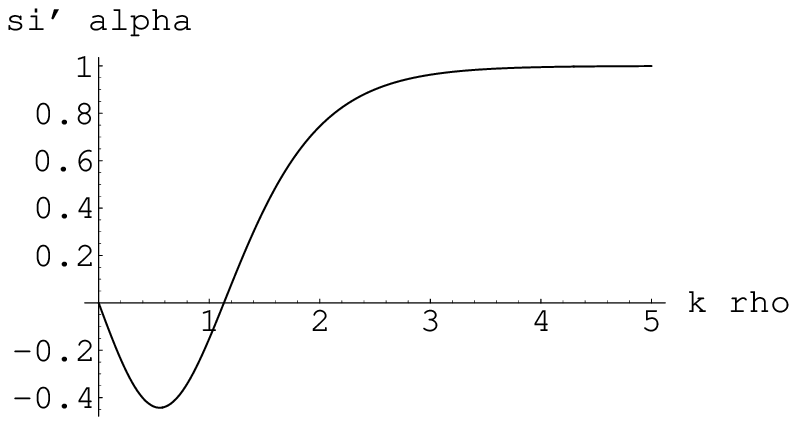}}
   \begin{center}
Fig.3\ The warp factor $\si'(\rho)/\al$,
(\ref{sol9}). Solution 2 in (\ref{vac2}) and (\ref{vac3}).
Horizontal axis: $k\rho$.
   \end{center}
\end{figure}

\begin{figure}
\centerline{\epsfysize=4cm\epsfbox{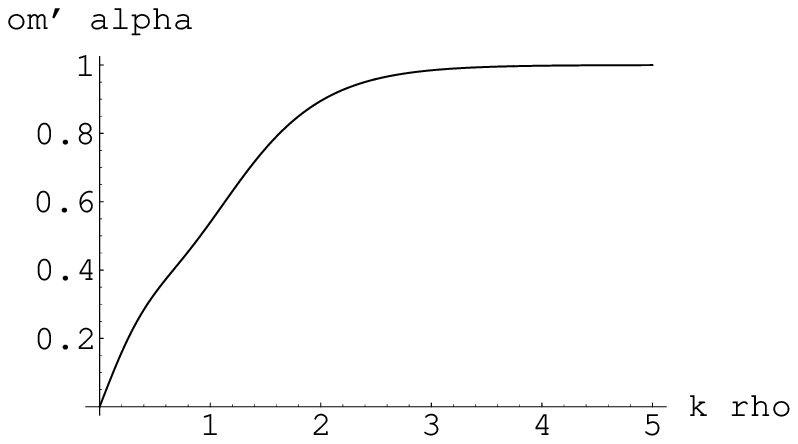}}
   \begin{center}
Fig.4\ The warp factor $\om'(\rho)/\al$,
(\ref{sol9}). Solution 2 in (\ref{vac2}) and (\ref{vac3}).
Horizontal axis: $k\rho$.
   \end{center}
\end{figure}

\begin{figure}
\centerline{\epsfysize=4cm\epsfbox{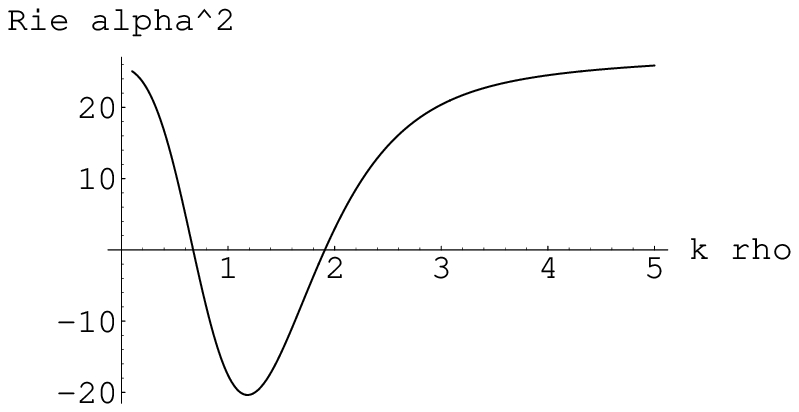}}
   \begin{center}
Fig.5\ 6D Riemann Scalar Curvature $\Rhat(\rho)/\al^2$,
(\ref{model3}). Solution 2 in (\ref{vac2}) and (\ref{vac3}).
Horizontal axis: $k\rho$.
   \end{center}
\end{figure}

\section{Numerical Solution}
Instead of the previous analytic approach based on the
assumed function form (\ref{sol9}), we can directly solve
the coupled differential equations Eq.(A), Eq.(B) and Eq.(C)
((\ref{sol5a}),(\ref{sol5b}) and (\ref{sol4c}))
in the numerical method (Runge-Kutta method). The difficult
point of this approach is the right choice of the vacuum
parameters. The three parameters should be correctly chosen
corresponding to a given boundary condition. 
(In the previous section, say, the condition is the initial
value of P:\ $P(\rho=0)/\vz= e_0$.) Another difficulty comes from
the technical reason in the numerical analysis. In the coupled
equation, there exists the "superficial" singularity at $\rho=0$. 
(Of course, there is no singularity in the obtained solution.)
To avoid the singular point, we must start, not from $\rho=0$ 
but from a point slightly away from $\rho=0$ in the positive
direction, say, $k\rho=0.1$. In this case, we must properly
choose the initial values of $P(k\rho=0.1), \si'(k\rho=0.1)$ and
$\om'(k\rho=0.1)$. For these values
(the vacuum parameters and the initial values)
we can borrow the information
from the analytic result in the previous section.
\footnote{
See App.B (i)Ultra-Violet Start for detail.
}
In Fig.6,7, and 8, we plot the results for
$P(\rho)$, $\si'(\rho)$ and $\om'(\rho)$ respectively.

\begin{figure}
\centerline{\epsfysize=4cm\epsfbox{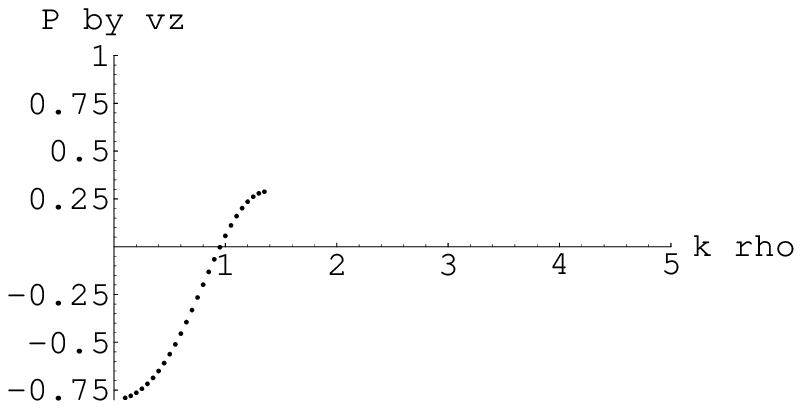}}
   \begin{center}
Fig.6\ The Higgs Field $P(\rho)/\vz$ obtained by the 
Runge-Kutta method. Ultra start. Horizontal axis: $k\rho$.
   \end{center}
\end{figure}

\begin{figure}
\centerline{\epsfysize=4cm\epsfbox{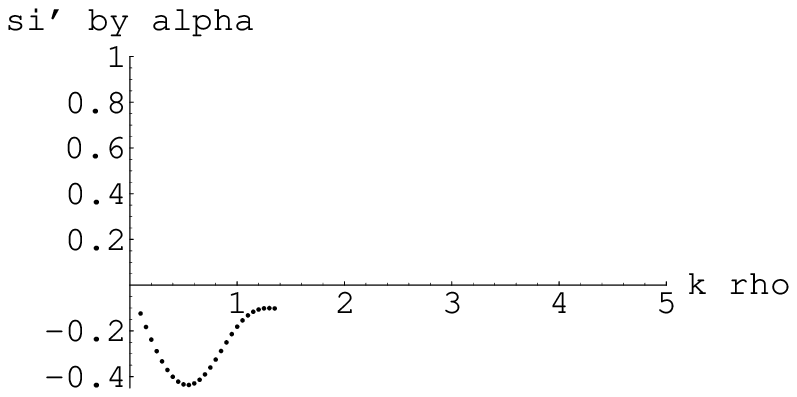}}
   \begin{center}
Fig.7\ The warp factor $\si'(\rho)/\al$ obtained by the 
Runge-Kutta method. Ultra start. Horizontal axis: $k\rho$.
   \end{center}
\end{figure}

\begin{figure}
\centerline{\epsfysize=4cm\epsfbox{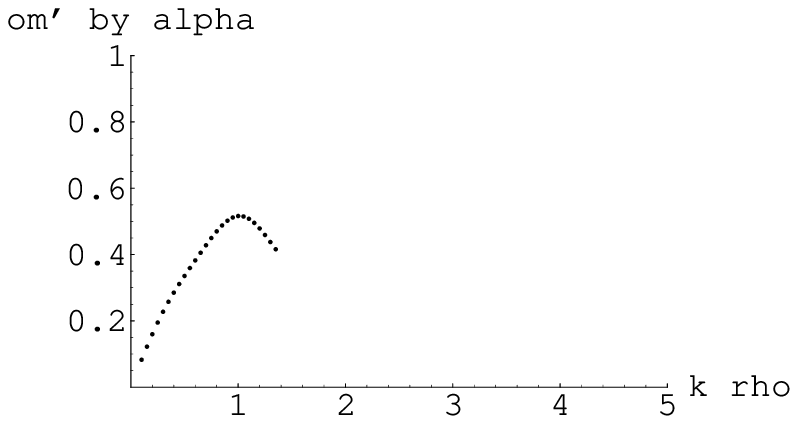}}
   \begin{center}
Fig.8\ The warp factor $\om'(\rho)/\al$ obtained by the 
Runge-Kutta method. Ultra start. Horizontal axis: $k\rho$.
   \end{center}
\end{figure}

The numerical results well reproduce the solution 
of the previous analytic approach in
the ultra-violet region. 
(Note that we do not assume the form of the solution (\ref{sol9}).)
The calculation, however, breaks down at $k\rho\approx 1$.
This is because the insufficiency of the correct choice of
the parameters and the initial values. 
We already know how finely those quantities should be
chosen, in the 5D case, in order to obtain
the required boundary condition\cite{SI0008}. 
See Appendix B
for detail, where the results of the infra-start are also
presented.

Note that the fact that the cosmological constant
and other parameters must be "fine-tuned" to give
the required boundary condition, just means those
are fixed to be some definite values. 
It is similar to the situation the energy eigen values
of the hydrogen atom are fixed by the boundary condition.
It does {\it never}
mean unstableness of the solution.

\section{Discussion and conclusion}
We summarize the present result by comparing it with the case
of 5D model by Randall and Sundrum.

\vs 1

\begin{tabular}{|c|c|c|}
\hline
  & 5 Dim.(Randall-Sundrum)            & 6 Dim.                            \\
\hline
Dimensions of & [$\vz$]=(mass)$^{3/2}$ & [$\vz$]=(mass)$^{2}$ \\
Vacuum & [$\la$]=(mass)$^{-1}$ & [$\la$]=(mass)$^{-2}$ \\
 Parameters     & [$\La$]=(mass)$^{5}$ & [$\La$]=(mass)$^{6}$ \\
\hline
Cond. on  & $-\frac{\la\vz^4}{4}<\La<0 $    &   $\La<0 $   \\
Cosm. Term        & Lower Bound exists        & No Lower Bound      \\
\hline
Cond. on        & $\frac{1}{r_c}\ll k\ll M$ &  $\frac{1}{\rho_c}\ll k\ll M$  \\
 Fund. Parameters    &  & \\
\hline
Asym. Behav.                     & $-\La\sim M^3k^2$ & $-\La\sim M^4k^2$  \\
of                              & $\vz\sim M^{3/2}$ &  $\vz\sim M^{2}$ \\
 Vac. Parameters               & $\la\sim M^{-3}k^2$ &  $\la\sim M^{-4}k^2$ \\
(same $k$-dep.)   & $\sqrt{\la\vz^2}\sim k$ & $\sqrt{\la\vz^2}\sim k$  \\
\hline
Mass Relations & $\frac{M_{pl}}{M}\sim \sqrt{\frac{M}{k}}$ 
                     & $\frac{M_{pl}}{M}\sim \frac{M}{k}$ \ \ see-saw rel.\\
  & $\La_{4d}= M_{pl}^2{\tilde \La}_{4d}\sim -M^3k$ 
               &  $\La_{4d}= M_{pl}^2{\tilde \La}_{4d}\sim -M^4$\\
\hline
4D Cosm. Term & $\sqrt{-{\tilde \La}_{4d}}\sim k\sim M^3\times 10^{-38}$GeV         & $\sqrt{-{\tilde \La}_{4d}}\sim k\sim M^2\times 10^{-19}$GeV  \\
\hline
\multicolumn{3}{c}{\q}                                                 \\
\multicolumn{3}{c}{Table 1\ \ Comparison of 
5D model and 6D model.  }\\
\end{tabular}

\vs 1

We add some numerical fact about
the see-saw relation (\ref{para3c}).
If we take as $k\sim 10^{-41}$GeV (inverse of the cosmological size),
we obtain $M\sim 10^{-11}$Gev$=10^{-2}$eV which is the order of the
neutrino mass. This is the case 1) in the second paragraph
of Subsec.5.3, which was abandoned there.
\footnote{
This choice means, as referred in subsec.5.3, the space-time
behaves as six dimensional at the cosmological scale. In this
connection, some interesting discussions are made in \cite{GRS002,Wit0002}. 
They examine some scenario where the world looks like D-dimensions
with D$\ge$5, above a very large (cosmological) size. 
}
This choice is, however, attractive in that 
it gives the right value of the cosmological constant. 
If this numerical fact is not accidental and has meaning,
it says the cosmological size is related to the neutrino mass
when it is "see-sawed" with the Planck mass.
These three fundamental scales are geometrically related.
The 6D fundamental scale $M$ gives the mass value,
and which suggests the neutrino mass is radiatively
generated in 6D {\it quantum} gravity.  
We consider this is an interesting mechanism
to explain the fact that, 
in the log-scale plot of the characteristic mass scales, 
the neutrino mass is located at the middle of the two scales: 
the Planck scale($10^{19}$GeV) and
the cosmological constant scale($10^{-41}$GeV).
Of course, in order to confidently accept this view, 
we should solve
the problem presented in Subsec.5.3. 

One of the fundamental parameters, $\rho_c$, 
is introduced in Subsec.5.1
as the infrared regularization.
This is quite natural in the standpoint of 
the discretized approach such as the lattice.
The treatment, however, should be regarded as 
an "effective" approach or a "temporary" stage 
of a right treatment.
One proposal is made in \cite{SI0008} for the
5D case where a new infrared regularization
is introduced. The metric (\ref{model2}) does not have
any singularity or horizon in the region $0\leq \rho <\infty$.
$\rho=\infty$ can be regarded as a horizon. The deficit angle
there is $2\pi$. 

We hope the present result of the hierarchy model will lead to
the bridge between the field theory of the soliton physics
and the string field theoy of the D-brane physics.

\vspace{2cm}

\vs 1
\begin{flushleft}
{\bf Acknowledgment}
\end{flushleft}
Part of the work was done during the stay at the summer institute SI2000
(Aug.7-14,2000),Fuji-Yoshida, Japan. The author thanks N. Ikeda for
discussions there and some calculational checks.
As for much advice on the numerical work, 
the author thanks T.Tamaribuchi.

\vs 1

\section{Appendix A: Calculation of Coefficients}
In this appendix we solve the REE ($M=1$) of Sec.3:
\begin{eqnarray}
\mbox{Eq.(A)}\q 3\si''-\frac{\si'}{\rho}+\si'\om'+\om''-(\om')^2
+2\frac{\om'}{\rho}
=2{P'}^2\com\mbox{(\ref{sol5a})}\nn
\mbox{Eq.(B)}\q -8{\si'}^2+2\frac{\si'}{\rho}-2\si'\om'+2\si''
=V\com\mbox{(\ref{sol5b})}\nn
\mbox{Eq.(C)}\q 4\si''-10{\si'}^2
=({P'}^2-m^2\frac{\e^{2\om}}{\rho^2}P^2+V)\com\mbox{(\ref{sol4c})}
\com\nonumber
\end{eqnarray}
in the following form:
\begin{eqnarray}
\si'(\rho)=\al\sum_{n=0}^\infty\frac{c_{2n+1}}{(2n+1)!}\{\tanh (k\rho)\}^{2n+1}\com\nn
\om'(\rho)=\al\sum_{n=0}^\infty\frac{d_{2n+1}}{(2n+1)!}\{\tanh (k\rho)\}^{2n+1}\com\nn
P(\rho)=v_0\sum_{n=0}^\infty\frac{e_{2n}}{(2n)!}\{\tanh (k\rho)\}^{2n}\com
\mbox{(\ref{sol9})}\nonumber
\end{eqnarray}
where 
$\al=\sqrt{-\La/10}M^{-2}$, (\ref{sol7}), and
$V=(\la/4)(P^2-\vz^2)^2+\La$, (\ref{model1}).
$\la,\vz$ and $\La$ are some constants called {\it vacuum
parameters} in the text. $1/k$ is another parameter called
{\it thickness} parameter. The integer parameter $m$ in Eq.(C) is
taken to be $m=0$ as explained in (\ref{sol7}). 
The coefficient-constants 
$c$'s, $d$'s and $e$'s are to be determined. 
$1/\rho$, appearing above, can be expressed as
\begin{eqnarray}
\frac{1}{\rho}=\frac{2k}{\tanh (k\rho)}
\sum_{n=0}^\infty\frac{s_{2n}}{(2n)!}\{\tanh (k\rho)\}^{2n}\com\nn
\mbox{where}\q
\left.\frac{d^{2n}}{dx^{2n}}\left(\frac{x}{\ln\frac{1+x}{1-x}}\right)\right|_{x=0}
\equiv s_{2n}\com\nn
s_0=\half\com\q s_2=-\third\com\q s_4=-\frac{12}{5}\com\q \cdots\com
\label{app3}
\end{eqnarray}
The above expansion formula is a key equation to solve the REE
in the form of (\ref{sol9}). 
All coefficients are finite as shown above. 
We can take the limit $\rho\ra\infty$ above, 
and see 
the infinite series {\it converges} and gives 0\ :\ 
$\sum_{n=0}^\infty\frac{s_{2n}}{(2n)!}=0$. 
It is a key, in the present treatment of the infinite series,  
that we take the expansion using 
powers of $tanh\,\rho$ not those of
$\e^{-2\rho}$.  
\footnote{
We can {\it not} reexpress the RHS of 
the first equation of (\ref{app3}) as
$1/\rho=\sum_{n=0}^\infty\frac{s'_{2n}}{(2n)!}\e^{-2n\rho}$ 
with {\it finite} coefficients.
The same notice is said about (\ref{sol9}).  
}

In the following of this appendix, we take the abbreviated notation
$t\equiv \tanh (k\rho)$. Using a relation
$t'=\frac{d}{d\rho}\tanh (k\rho)=k(1-t^2)$, we obtain
\begin{eqnarray}
P'=k\vz (1-t^2)\sum_{n=1}^\infty\frac{e_{2n}}{(2n-1)!}\,t^{2n-1}\com\nn
\si''=k\al (1-t^2)\sum_{n=0}^\infty\frac{c_{2n+1}}{(2n)!}\,t^{2n}
=k\al \{ c_1+\sum_{n=1}^\infty (\frac{c_{2n+1}}{(2n)!}-\frac{c_{2n-1}}{(2n-2)!})\,t^{2n}\}
\com\nn
\om''=k\al (1-t^2)\sum_{n=0}^\infty\frac{d_{2n+1}}{(2n)!}\,t^{2n}
=k\al \{ d_1+\sum_{n=1}^\infty (\frac{d_{2n+1}}{(2n)!}-\frac{d_{2n-1}}{(2n-2)!})\,t^{2n}\}
\pr\label{app3b}
\end{eqnarray}

Several useful expansion formulae are listed below.
\begin{eqnarray}
\frac{\si'}{\rho}= 2k\al
\sum_{N=0}^\infty[sc]_N\,t^{2N}\q\mbox{where}\q
[sc]_N\equiv \sum_{n=0}^N
\frac{s_{2n}}{(2n)!}\frac{c_{2N-2n+1}}{(2N-2n+1)!}\com\nn
\ [sc]_0=s_0c_1=\half c_1\ ,\ [sc]_1=s_0\frac{c_3}{3!}+\frac{s_2}{2!}c_1
=\frac{c_3}{12}-\frac{c_1}{6}\ ,
\ [sc]_2=\frac{c_5}{240}-\frac{c_3}{36}-\frac{c_1}{10}\ ,\  
\cdots \nn
\frac{\om'}{\rho}= 2k\al
\sum_{N=0}^\infty[sd]_N\,t^{2N}\q\mbox{where}\q
[sd]_N\equiv \sum_{n=0}^N
\frac{s_{2n}}{(2n)!}\frac{d_{2N-2n+1}}{(2N-2n+1)!}\com\nn
\om'\si'= \al^2
\sum_{N=0}^\infty[dc]_N\,t^{2N+2}\q\mbox{where}\q
[dc]_N\equiv \sum_{n=0}^N
\frac{d_{2n+1}}{(2n+1)!}\frac{c_{2N-2n+1}}{(2N-2n+1)!}\com\nn
\ [dc]_0=d_1c_1\ ,\ [dc]_1=\frac{1}{6}(d_1c_3+d_3c_1)\ ,\ \cdots \nn
{\om'}^2= \al^2
\sum_{N=0}^\infty[dd]_N\,t^{2N+2}\q\mbox{where}\q
[dd]_N\equiv \sum_{n=0}^N
\frac{d_{2n+1}}{(2n+1)!}\frac{d_{2N-2n+1}}{(2N-2n+1)!}\com\nn
{\si'}^2= \al^2
\sum_{N=0}^\infty[cc]_N\,t^{2N+2}\q\mbox{where}\q
[cc]_N\equiv \sum_{n=0}^N
\frac{c_{2n+1}}{(2n+1)!}\frac{c_{2N-2n+1}}{(2N-2n+1)!}\com\nn
\ [cc]_{0}=(c_1)^2\com\q [cc]_1=\third c_1c_3\com\q\cdots\com\nn
{P}^2= \vz^2
\sum_{N=0}^\infty[ee]_N\,t^{2N}\q\mbox{where}\q
[ee]_N\equiv \sum_{n=0}^N
\frac{e_{2n}}{(2n)!}\frac{e_{2N-2n}}{(2N-2n)!}\com\nn
\ [ee]_{0}=(e_0)^2\com\q [ee]_1=e_0e_2\com\q [ee]_2=\frac{1}{12}e_0e_4+\fourth (e_2)^2\com
\cdots\com\nn
{P'}^2= k^2\vz^2(1-t^2)^2
\sum_{N=1}^\infty[ee]'_N\,t^{2N}\nn
=k^2\vz^2\{ [ee]'_1t^2+([ee]'_2-2[ee]'_1)t^4+
\sum_{N=3}^\infty ([ee]'_N-2[ee]'_{N-1}+[ee]'_{N-2})t^{2N} \}\com\nn
\q\mbox{where}\q
[ee]'_N\equiv \sum_{n=1}^N
\frac{e_{2n}}{(2n-1)!}\frac{e_{2N-2n+2}}{(2N-2n+1)!}\com\nn
\ [ee]'_1=(e_2)^2\com\q [ee]'_2=\third e_2e_4\com\cdots\nn
(P^2)^2= \vz^4
\sum_{N=0}^\infty[e^4]_N\,t^{2N}\q\mbox{where}\q
[e^4]_N\equiv \sum_{n=0}^N
[ee]_n[ee]_{N-n}\com\nn
\ [e^4]_{0}=(e_0)^4\com\q [e^4]_1=2(e_0)^3e_2\com\q 
[e^4]_2=\frac{1}{6}(e_0)^3e_4+\frac{3}{2}(e_0)^2(e_2)^2\com\cdots
\label{app4}
\end{eqnarray}
Now we find recursion relations between coefficients by inserting
expansion formulae above into the REE (\ref{sol5a},\ref{sol5b},\ref{sol4c}).

(A)\ From (\ref{sol5a}) we obtain
\begin{eqnarray}
t^0\mbox{-part}\q (c_1,d_1)\qqqq\qqq\qqq\nn
k\al\{ 2c_1+3d_1\}=0\com\label{app5a}\\
t^2\mbox{-part}\q (c_3,d_3,e_2)\qqqq\qqq\qqq\nn
k\al\{ \frac{4}{3}c_3+\frac{5}{6}d_3\}-2k^2\vz^2(e_2)^2=
k\al\{ \frac{8}{3}c_1+\frac{5}{3}d_1\}-\al^2(c_1-d_1)d_1\ ,\label{app5b}\\
t^4\mbox{-part}\q (c_5,d_5,e_4)\qqqq\qqq\qqq\nn
k\al\{ \frac{7}{60}c_5+\frac{7}{120}d_5 \}-\frac{2}{3}k^2\vz^2e_2e_4=
k\al\{ \frac{13}{9}c_3-\frac{1}{5}c_1           \nn
+\frac{11}{18}d_3+\frac{2}{5}d_1 \}
-\al^2\{\frac{1}{6}(d_1c_3+d_3c_1)-\frac{1}{3}d_1d_3 \}
-4k^2\vz^2(e_2)^2\ ,\label{app5c}\\
t^{2N}(N\ge 3)\mbox{-part}\q (c_{2N+1},d_{2N+1},e_{2N})\qqqq\qqq\qqq\nn
k\al\left\{ 3\left(\frac{c_{2N+1}}{(2N)!}-\frac{c_{2N-1}}{(2N-2)!}\right)
-2[sc]_N
+\left(\frac{d_{2N+1}}{(2N)!}-\frac{d_{2N-1}}{(2N-2)!}\right)
+4[sd]_N \right\}\nn
+\al^2\{ [dc]_{N-1}-[dd]_{N-1} \}
-2k^2\vz^2\{ [ee]'_N-2[ee]'_{N-1}+[ee]'_{N-2} \}=0\ ,
\label{app5d}
\end{eqnarray}

(B)\ From (\ref{sol5b}) we obtain
\begin{eqnarray}
t^0\mbox{-part}\q (c_1,e_0)\qqqq\qqq\qqq\nn
2k\al c_1-\frac{1}{8}\la\vz^4(1-(e_0)^2)^2=\half\La\com\label{app6a}\\
t^{2N}(N\ge 1)\mbox{-part}\q (c_{2N+1},e_{2N})\qqqq\qqq\qqq\nn
k\al\left\{ (\frac{c_{2N+1}}{(2N)!}-\frac{c_{2N-1}}{(2N-2)!})+2[sc]_N \right\}
-\al^2\{ [dc]_{N-1}+4[cc]_{N-1} \}     \nn
+\la\vz^4\{ \fourth [ee]_N-\frac{1}{8}[e^4]_N \}=0\ ,
\label{app6b}
\end{eqnarray}

(C)\ From (\ref{sol4c}) we obtain
\begin{eqnarray}
t^0\mbox{-part}\q (c_1,e_0)\qqqq\qqq\qqq\nn
4k\al c_1-\fourth\la\vz^4(1-\ez^2)^2=\La\com\label{app7a}\\
t^2\mbox{-part}\q (c_3,e_2)\qqqq\qqq\qqq\nn
2k\al c_3-k^2\vz^2(e_2)^2+\half\la\vz^4\ez(1-\ez^2)e_2=
4k\al c_1+10\al^2(c_1)^2\ ,\label{app7b}\\
t^4\mbox{-part}\q (c_5,e_4)\qqqq\qqq\qqq\nn
\frac{1}{6}k\al c_5
+\left\{ -\frac{1}{3}k^2\vz^2e_2+\frac{1}{24}\la\vz^4\ez(1-\ez^2) \right\}e_4=\nn
2k\al c_3+\frac{10}{3}\al^2 c_1c_3
-2k^2\vz^2(e_2)^2-\frac{1}{8}\la\vz^4(1-3\ez^2)(e_2)^2\ ,\label{app7c}\\
t^{2N}(N\ge 3)\mbox{-part}\q (c_{2N+1},e_{2N})\qqqq\qqq\qqq\nn
4k\al \left( \frac{c_{2N+1}}{(2N)!}-\frac{c_{2N-1}}{(2N-2)!}\right)
-10\al^2 [cc]_{N-1}
-k^2\vz^2\{ [ee]'_N-2[ee]'_{N-1}+[ee]'_{N-2} \}\nn
+\fourth\la\vz^4 (2[ee]_N-[e^4]_N)
=0\ ,
\label{app7d}
\end{eqnarray}
We have now obtained all necessary recursion relations and 
are now ready for determining all coefficients $(c_{2n+1},d_{2n+1},e_{2n})$.\nl

i)\ $t^0$-part,\ $(c_1,d_1,e_0)$\nl
\q Relevant equations are (\ref{app5a}),(\ref{app6a}) and (\ref{app7a}).
(\ref{app6a}) is, however, equivalent to (\ref{app7a}). 
Hence one coefficient, say $e_0$, remains as a free parameter.
\begin{eqnarray}
c_1=\frac{M^{-4}}{4k\al}\{ \fourth\la\vz^4(1-\ez^2)^2+\La \}\com\nn
d_1=-\frac{2}{3}c_1\com\nn
e_0=\frac{1}{\vz}P(\rho=0)\ :\ \mbox{free parameter}\com
\label{app8}
\end{eqnarray}

ii)\ $t^2$-part,\ $(c_3,d_3,e_2)$\nl
\q Relevant equations are (\ref{app5b}),(\ref{app6b}) with $N=1$ and (\ref{app7b}).
There are two solutions.
\begin{eqnarray}
\mbox{Solution 1}\qqq\qqq\qqq\nn
e_2=0\com\q
c_3=2c_1+5\frac{\al}{k}(c_1)^2\com\q
d_3=-\frac{4}{3}(c_1+5\frac{\al}{k}(c_1)^2)\nn
\mbox{Solution 2}\qqq\qqq\qqq\nn
e_2=-\fourth\frac{\la\vz^2}{k^2}\ez(1-\ez^2)\ ,\ \ez\neq 0,\pm 1,\nn
c_3=\frac{3}{32}\frac{\la^2\vz^6}{k^3\al}M^{-4}
\ez^2(1-\ez^2)^2+2c_1+5\frac{\al}{k}(c_1)^2\ ,\nn
d_3=-\frac{4}{3}(c_1+5\frac{\al}{k}(c_1)^2)
\pr
\label{app9}
\end{eqnarray}
Solution 2 includes Solution 1 if we allow $\ez= 0,\pm 1$.  

iii)\ $t^4$-part,\ $(c_5,d_5,e_4)$\nl
\q Relevant equations are (\ref{app5c}),(\ref{app6b}) with $N=2$ and (\ref{app7c}).
They are solved as
\begin{eqnarray}
\{k^2\vz^2e_2+\frac{1}{12}\la\vz^4\ez(1-\ez^2)\}e_4\nn
=k\al (-\frac{4}{9}c_3+2c_1)
+\al^2\cdot\frac{5}{3}(d_1c_3+d_3c_1+2c_1c_3)\nn
-\la\vz^4\fourth (1-3\ez^2){e_2}^2+6k^2\vz^2{e_2}^2\com\nn
\al c_5=-\frac{5}{12}\frac{\la\vz^4}{k}\ez(1-\ez^2)e_4
+\frac{100}{9}\al c_3+4\al c_1        \nn
+\frac{\al^2}{k}\cdot\frac{10}{3}(d_1c_3+d_3c_1+8c_1c_3)
-\frac{5}{4}\frac{\la\vz^4}{k}(1-3\ez^2){e_2}^2\com\nn
\al d_5=-2\al c_5+\frac{120}{7}\{
\frac{2}{3}k\vz^2e_2e_4+\frac{13}{9}\al c_3\nn
-\frac{1}{5}\al c_1+\frac{11}{18}\al d_3+\frac{2}{5}\al d_1\nn
-\frac{\al^2}{k}\cdot\frac{1}{6}(d_1c_3+d_3c_1-2d_1d_3)
-4k\vz^2{e_2}^2
                              \}
\label{app10}
\end{eqnarray}

iv)\ $t^{2N}(N\geq 3)$-part,\ $(c_{2N+1},d_{2N+1},e_{2N})$\nl
\q Relevant equations are (\ref{app5d}),(\ref{app6b}) and (\ref{app7d}).
\begin{eqnarray}
k\al\left\{ 3\left(\frac{c_{2N+1}}{(2N)!}-\frac{c_{2N-1}}{(2N-2)!}\right)
-2[sc]_N
+\left(\frac{d_{2N+1}}{(2N)!}-\frac{d_{2N-1}}{(2N-2)!}\right)
+4[sd]_N \right\}\nn
+\al^2\{ [dc]_{N-1}-[dd]_{N-1} \}
-2k^2\vz^2\{ [ee]'_N-2[ee]'_{N-1}+[ee]'_{N-2} \}=0\ ,\nn
k\al\left\{ (\frac{c_{2N+1}}{(2N)!}-\frac{c_{2N-1}}{(2N-2)!})+2[sc]_N \right\}
-\al^2\{ [dc]_{N-1}+4[cc]_{N-1} \}     \nn
+\la\vz^4\{ \fourth [ee]_N-\frac{1}{8}[e^4]_N \}=0\ ,\nn
4k\al \left( \frac{c_{2N+1}}{(2N)!}-\frac{c_{2N-1}}{(2N-2)!}\right)
-10\al^2 [cc]_{N-1}
-k^2\vz^2\{ [ee]'_N-2[ee]'_{N-1}+[ee]'_{N-2} \}\nn
+\fourth\la\vz^4 (2[ee]_N-[e^4]_N)
=0\ ,
\label{app11}
\end{eqnarray}
Because $[sc]_N,[sd]_N,[ee]'_N,[ee]_N$ and $[e^4]_N$ have
the terms $c_{2N+1},d_{2N+1}$ and $e_{2N}$ {\it linearly} (for $N\geq 3$) and
other parts of the relevant equations are made of lower-order terms,
they are three coupled {\it linear} equations for the three
variables $c_{2N+1},d_{2N+1}$ and $e_{2N}$. Hence  
we conclude that $(c_{2N+1},d_{2N+1},e_{2N})$ are fixed
by the lower-order coefficients.

\section{Appendix B: Numerical Analysis by Runge-Kutta}
We explain the numerical analysis of Sec.7 of the text.
Instead of the functions $\si'$,$\om'$ and $P$, it is
convenient to use normalized ones:\ 
\begin{eqnarray}
\Ptil=\frac{P}{\vz}\com\q
\Si=\frac{\si'}{\al}\com\q
\Om=\frac{\om'}{\al}\ .
\label{appB1}
\end{eqnarray}
They satisfy the unit boundary condition:\ 
$\Ptil, \Si, \Om \ra 1$ as $\rho\ra \infty$. In terms of
these functions, we can rewrite the REE equations (\ref{sol5a},
\ref{sol5b},\ref{sol4c}) as
\begin{eqnarray}
\frac{1}{k}\frac{d}{d\rho}\Si=-\frac{1}{k\rho}\Si+
\frac{\al}{k}(\Si\Om+4\Si^2)
+\frac{1}{8}\frac{\la\vz^4M^{-4}}{k\al}(\Ptil^2-1)^2
+\half\frac{\La M^{-4}}{k\al}\com\nn
\frac{1}{k}\frac{d}{d\rho}\Om=-\frac{4}{k\rho}\Si-\frac{2}{k\rho}\Om
+\frac{\al}{k}(4\Si\Om+\Om^2)
+\frac{1}{8}\frac{\la\vz^4M^{-4}}{k\al}(\Ptil^2-1)^2
+\half\frac{\La M^{-4}}{k\al}\com\nn
\frac{1}{k}\frac{d}{d\rho}\Ptil=+
\left(
-4\frac{\al M^4}{k\vz^2}\frac{1}{k\rho}\Si
+4\frac{\al^2 M^4}{k^2\vz^2}\Si\Om
\right.  \nn
\left.
+6\frac{\al^2 M^4}{k^2\vz^2}\Si^2
+\frac{1}{4}\frac{\la\vz^2}{k^2}(\Ptil^2-1)^2
+\frac{\La}{k^2\vz^2}
\right)^{1/2}\pr
\label{appB2}
\end{eqnarray}
This is the standard form of the Runge-Kutta equation
for $\Ptil,\Si$ and $\Om$ with the coordinate $k\rho$. 
All dimensionless
constants appearing in the RHSs are rewritten, in terms of
the dimensionless vacuum parameters in (\ref{vac1}):\ 
$\sqrt{-\La}/M^2k\equiv A,\ 
\la\vz^4/M^4k^2\equiv L,\ 
\vz^2/M^4\equiv B$, as
\begin{eqnarray}
\frac{-\La M^{-4}}{k\al}=\sqrt{10}A\com\q
\frac{\la\vz^4M^{-4}}{k\al}=\sqrt{10}\frac{L}{A}\com\q
\frac{\al}{k}=\frac{A}{\sqrt{10}}\com\nn
\frac{\al M^4}{k\vz^2}=\frac{1}{\sqrt{10}}\frac{A}{B}\com\q
\frac{\la\vz^2}{k^2}=\frac{L}{B}\com\q
\frac{-\La}{k^2\vz^2}=\frac{A^2}{B}\com\nn
\frac{\al^2 M^4}{k^2\vz^2}=\frac{1}{10}\frac{A^2}{B}
\pr
\label{appB3}
\end{eqnarray}
The values of $(A,B,L)$ must be properly obtained
corresponding to a given boundary condition.
For the case $\Ptil(k\rho=0)=e_0=-0.8$ case (Solution 2), 
we have obtained
their values as $(A,B,L)\sim (\al_0,\be_0,\del_0)=  \nl
(1.864786, 0.6424929, 15.90691)$ in Sec.6. We take these values
in this appendix.
As for the initial condition for ($\Ptil, \Si, \Om$), 
we analyze two cases.

(i) Ultra-Violet Start\nl
First we numerically solve the differential equation
in the direction 
from the small $k\rho$ value to the large one.
Because the origin $k\rho=0$ is the "superficial" singularity
in (\ref{appB2}), we must take, as the initial point, 
some point slightly away (in the positive direction) from
the origin. Say, $k\rho=0.1$. As the boundary value
of $\Ptil(0.1),\Si(0.1)$ and $\Om(0.1)$, we take the values
obtained from the analytic result of Sec.6. They are
-0.791057,-0.123938 and 0.082700, respectively. The results are
shown in Fig.6,7,and 8 of Sec.7. Those figures surely
reproduce the analytical result of Sec.6 up to $k\rho\sim 1$.
It strongly supports the correctness of the solution
found in Sec.3 and 6. It also means the validity of
the truncation approximation to determine the
vacuum parameters. 
The calculation stops with producing a imaginary value.
It occurs at the place where the inside of
the square root in the last equation of (\ref{appB2})
approaches 0 but is still required to be non-negative. 
The present precision is insufficient to keep non-negative
in such nearly 0 region. 

(ii) Infra-Red Start\nl
Next we numerically solve the equations 
in the direction from the large
$k\rho$ value to the small one. Say, from $k\rho=3.1$
to $k\rho=0.1$. As for the initial values
$\Ptil(3.1),\Si(3.1)$ and $\Om(3.1)$, one idea is to
take the analytic result of Sec.6, as in the (i) case. 
It, however, fails.
The more stringent requirement of the positivity, 
than the case of the UV start case, is demanded. 
We can see the situation by writing the inside of
the square root of (\ref{appB2}) as
\begin{eqnarray}
\left(\frac{1}{k}\frac{d}{d\rho}\Ptil\right)^2=
-\frac{4}{\sqrt{10}}\frac{A}{B}\frac{1}{k\rho}\Si
+\frac{1}{5}\frac{A^2}{B}(3\Si+5\Om)(\Si-\Om)\nn
-\frac{A^2}{B}(1-\Om^2)+\frac{1}{4}\frac{L}{B}(\Ptil^2-1)^2
\ge 0
\pr
\label{appB4}
\end{eqnarray}
In the asymptotic region $k\rho\ra \infty$, we know
$\Ptil\ra 1,\ \Si\ra 1,\ \Om\ra 1,\ \Si/k\rho\ra 0$. 
Therefore the above quantity becomes quite small
and, at the same time, is required to be non-negative.
The failure means 
the approximation of the 3-terms truncation
used in Sec.6 is insufficient in the IR region. 
Here we cannot help being
content with the qualitative correctness
of the solution. In Fig.9,10,11, we show the results
for the initial data:\ the initial point $k\rho=3.1$;\ 
the initial values 
$\Ptil(3.1)=1.0,\Si(3.1)=1.12,\Om(3.1)=1.12$. 
They are consistent with the analytical result
in the qualitative way.

\begin{figure}
\centerline{\epsfysize=4cm\epsfbox{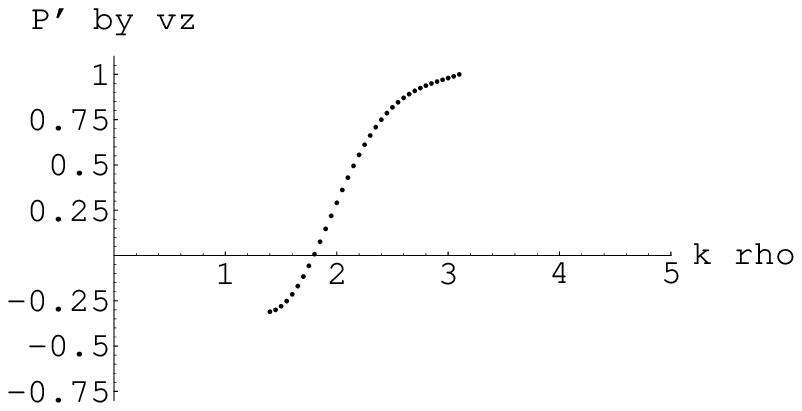}}
   \begin{center}
Fig.9\ The Higgs Field $P(\rho)/\vz$ obtained by the 
Runge-Kutta method. Infra start. Horizontal axis: $k\rho$.
   \end{center}
\end{figure}

\begin{figure}
\centerline{\epsfysize=4cm\epsfbox{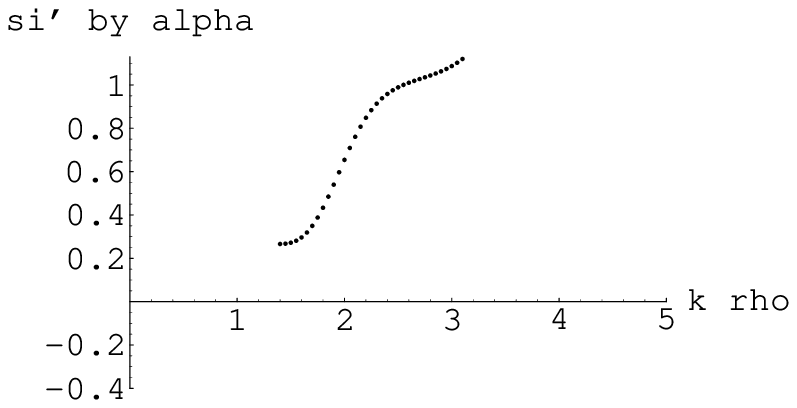}}
   \begin{center}
Fig.10\ The warp factor $\si'(\rho)/\al$ obtained by the 
Runge-Kutta method. Infra start. Horizontal axis: $k\rho$.
   \end{center}
\end{figure}

\begin{figure}
\centerline{\epsfysize=4cm\epsfbox{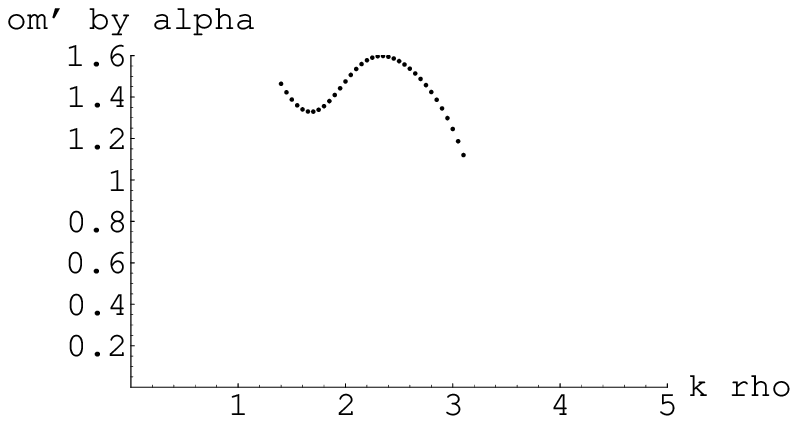}}
   \begin{center}
Fig.11\ The warp factor $\om'(\rho)/\al$ obtained by the 
Runge-Kutta method. Infra start. Horizontal axis: $k\rho$.
   \end{center}
\end{figure}



\begin{thebibliography}{99}
\bibitem{foot5a} 
Use of $\La_{4d}$ or ${\tilde \La}_{4d}$ is purely a convension. 
$\int d^4x\sqrt{-g}\La_{4d}=(1/G)\int d^4x\sqrt{-g}{\tilde \La}_{4d}$
where $1/G=M_{pl}^2$ ($G$: Newton constant).
\bibitem{foot5b} 
This was shown in \cite{SI95} for the 2 dim $R^2$ gravity
which is renormalizable.
\end{thebibliography}
\end{document}